\documentclass[aps,twocolumn,showpacs,pre]{revtex4-1}

\usepackage{graphicx}
\usepackage{amsmath,amssymb,amsfonts}
\usepackage{bbm,bbold}
\usepackage{multirow}
\usepackage{color}
\usepackage{placeins}
\usepackage{braket}

\usepackage[
    colorlinks=true,
    linkcolor=blue,
    citecolor=blue,
    urlcolor=blue
]{hyperref}

\newcommand{\rpsivec}[1]{\ensuremath{\left|\psi_{\vec{#1}}\right\rangle}}
\newcommand{\lpsivec}[1]{\ensuremath{\left\langle\bar{\psi}_{\vec{#1}}\right|}}
\newcommand{\rphivec}[1]{\ensuremath{\left|\phi_{\vec{#1}}\right\rangle}}
\newcommand{\lphivec}[1]{\ensuremath{\left\langle\bar{\phi}_{\vec{#1}}\right|}}

\begin{document}
\title[Encounter times of random walkers with simultaneous resetting on networks]{Encounter times of random walkers with simultaneous resetting on networks}
	\author{Cristian D. Suarez-Jimenez$^1$, Alejandro P. Riascos$^1$
		, and Denis Boyer${}^{2}$}  
	\address{		${}^1$Departamento de Física, Universidad Nacional de Colombia, Bogotá, Colombia\\
	${}^2$Instituto de F\'isica, Universidad Nacional Aut\'onoma de M\'exico, 
	C.P. 04510, Ciudad de M\'exico, M\'exico}
\begin{abstract}
In this work, we study the dynamics of multiple random walkers on networks subject to a simultaneous resetting protocol, whereby all walkers are synchronously returned to their respective initial nodes. For this collective Markovian process, we derive exact analytical expressions for the mean first-encounter time, defined as the average time required for all walkers to meet for the first time at a given node. These results are formulated in terms of the eigenvalues and eigenvectors of the transition matrices governing the dynamics without resetting, providing a clear spectral interpretation of the impact of resetting on encounter processes. We further establish a general criterion for finite networks that determines when the introduction of a nonzero resetting probability reduces the mean first-encounter time and leads to an optimal resetting strategy. The theoretical predictions are illustrated through numerical results on regular and heterogeneous networks, for encounters involving two or more walkers, and for combinations of local and nonlocal dynamics. Our findings demonstrate that simultaneous resetting can significantly reduce encounter times for specific target nodes and initial conditions, while becoming ineffective for highly exploratory dynamics or distant targets. A comparison with independent resetting shows that simultaneous resetting is more efficient in homogeneous networks, whereas independent resetting can outperform it in heterogeneous structures, thereby revealing a trade-off between synchronization and exploration. The framework provides a unified approach to collective search and encounter problems on networks with resetting.
\end{abstract}


\maketitle

\section{Introduction}
Diffusive transport and random-walk strategies have been employed for a long time across a wide range of disciplines as effective mechanisms for locating hidden targets or exploring spatial domains. Within this context, recent years have seen increasing interest in stochastic search processes that incorporate resetting or restart mechanisms. When a random process is intermittently interrupted and restarted stochastically to a prescribed configuration (typically its initial state), the statistical properties of the dynamics are significantly altered. In particular, the average time required to reach a specified target, known as the mean first-passage time, can often be minimized by tuning the resetting rate to a non-trivial value \cite{evans2011diffusion,Evans2011JPhysA,reuveni2016optimal,EvansReview2019}. A variety of resetting protocols and restart mechanisms have been proposed and analyzed over the past decade, considerably broadening the scope of stochastic resetting processes \cite{pal2016diffusion,nagar2016diffusion,Bhat2016JStat,chechkin2018random,PalPRE_2019,Nagar2023,SalgadoPRE_2024}, and have been applied to diverse classes of stochastic processes, including Brownian motion \cite{evans2011diffusion,Evans2011JPhysA,MajumdarPRE2015}, biased diffusive dynamics \cite{montero2013monotonic,ray2019peclet}, Lévy flights and other non-Gaussian search strategies \cite{kusmierz2015optimal,Kusmierz2014PRL,maso2019transport}, and models of anomalous diffusion \cite{Kusmierz2014PRL,kusmierz2015optimal,kusmierz2019subdiffusive}.
\\[2mm]
Moreover, a wide range of phenomena can be effectively modeled as dynamical processes on networks \cite{NewmanBook,barabasi2016book,VespiBook}. The interplay between the topology of networks and the dynamics taking place on them is central to understanding the behavior of many complex systems \cite{NewmanBook,VespiBook,VanMieghem2011}. In this context, random-walk strategies, where transitions only occur locally to an adjacent node with equal probability (the so-called normal random walk), provide a natural and widely applicable framework for analyzing diffusive transport on networks \cite{VespiBook,Hughes,Lovasz1996,MulkenPR502}. Significant progress has been made in understanding network exploration via random walks \cite{NohRieger2004,Tejedor2009PRE,MasudaPhysRep2017}. Despite these advances, random walks on networks subject to resetting have received comparatively less attention \cite{Avrachenkov2014,Avrachenkov2018,Touchette_PRE2018,ResetNetworks_PRE2020,christophorov2020peculiarities,Wald_PRE2021,bonomo2021first}. Recently, some connections were established between random-walk dynamics with resetting to a single node and the spectral decomposition of the transition matrix that defines the normal random walk 
\cite{Touchette_PRE2018,ResetNetworks_PRE2020,MultipleResetPRE_2021,Li_PRE_2025}. These insights underscore the potential of resetting mechanisms as efficient strategies for exploring diverse network topologies \cite{ResetNetworks_PRE2020,MultipleResetPRE_2021,riascos_jpa_2022,Michelitsch2025}.
\\[2mm]
Most of the aforementioned studies focus on the dynamics of a single random walker, while the motion of multiple walkers has received comparatively limited focus \cite{WengPRE2017}. Nevertheless, the presence of multiple agents is a common feature in real-world processes occurring on complex systems; for instance, in encounter networks arising from human activity \cite{RiascosMateosPlosOne2017,Mastrandrea_PlosOne2015}, epidemic spreading \cite{SatorrasPRL2001,ValdezBraunsteinHavlin2020,Bestehorn2021,Giuggioli_2022,Granger2024}, ecological interactions \cite{GiuggioliPRL2013,CraftRoyalB_2015}, and the emergence of extreme events \cite{KishorePRL2011}, among others. Recent efforts examining the efficiency of multiple searchers in locating target sites on networks include the study of the mean time required for one or more walkers to reach a given node for the first time \cite{DaiPhysA2020}, the emergence of universal laws governing search times \cite{WengPRE2017,WengPRE2018universal,DaiPhysA2020}, analytical descriptions of encounter times among many random walkers \cite{DPSandersPRE2009,riascos_multiple2020}, and search strategies involving the capture of moving targets with predefined trajectories \cite{Weng_2017,WengChaos2018}. 
\\[2mm]
Other recent studies underscore the relevance of resetting mechanisms in systems with multiple agents, for instance through independent individual resets \cite{Rubio-Gomez_2025}. A different scenario arises when a collection of searchers follows the same resetting protocol simultaneously. Examples include Brownian particles undergoing stochastic simultaneous resetting to a fixed position~\cite{BiroliPRL_2023,BiroliPRE_2023,biroli2024exact,deMauro_PRE_2026,Boyer_2026}, or subject to a threshold-triggered simultaneous restart~\cite{biswas2025_prl}.  More generally, interacting and many-particle systems exhibit a rich variety of resetting mechanisms, including global and local resetting protocols, subsystem restarts, and more general group-resetting strategies. These systems have recently attracted considerable attention in studies of stochastic resetting, search processes, and first-passage dynamics \cite{Falcao_2017,Nagar2023,Singh2023PRE,Singh2025Chaos,Lee2026PRR,Lee2026arXiv}, highlighting that the statistical properties and search efficiency can depend sensitively on the resetting protocol. In particular, simultaneity produces all-to-all correlations between the particles that persist asymptotically and can enhance performance in optimization and search tasks. Systems of non-interacting particles with simultaneous resetting belong to a more general class of processes where correlations emerge dynamically from a common fluctuating environment, as reviewed in Ref.~\cite{majumdar2026dynamically}.
\\[2mm]
\begin{figure}[!t]
	\begin{center}
		\includegraphics*[width=0.5\textwidth]{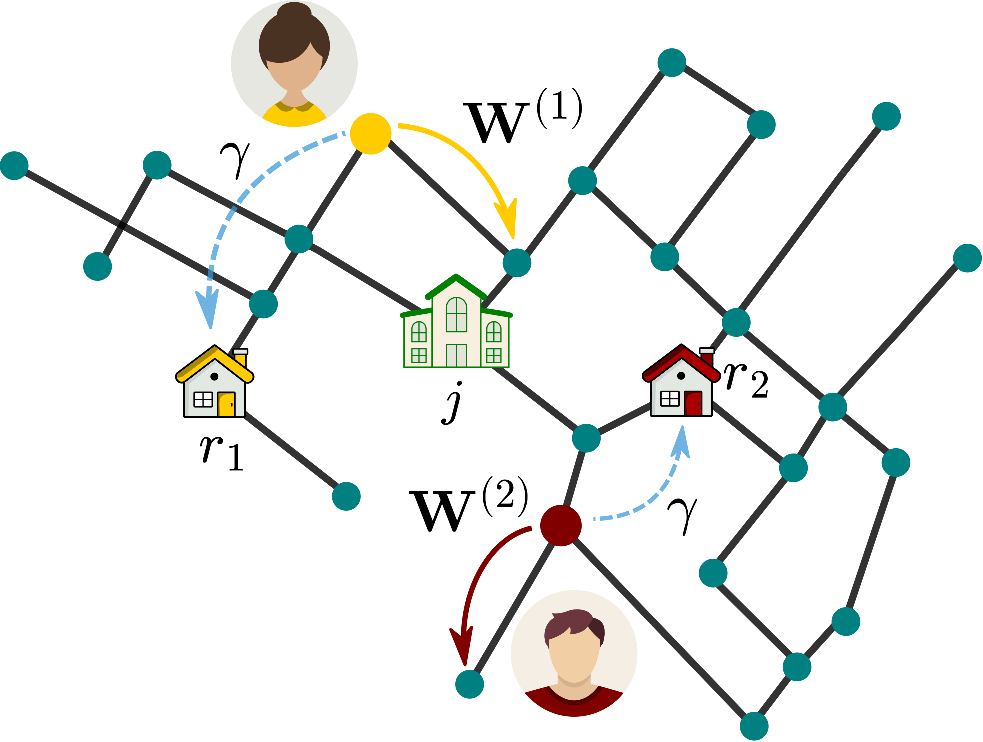}
	\end{center}
	\vspace{-5mm}
	\caption{\label{Fig_1} Two walkers move on a street network whose nodes represent points of interest in a city. Alice randomly moves according to the underlying transition matrix $\mathbf{W}^{(1)}$, while Bob follows a different dynamics defined by $\mathbf{W}^{(2)}$. With probability $\gamma$, both walkers reset simultaneously to their respective home nodes $r_1$ and $r_2$ (indicated by the dashed arrows), whereas with probability $1-\gamma$ they continue their random-walk motion. The goal is to compute the mean time until both walkers meet at node $j$.}
\end{figure}
In this paper, we develop a general framework to study the collective dynamics of multiple non-interacting random walkers subject to simultaneous resetting on arbitrary networks, where each walker evolves according to its own transition matrix. We derive exact analytical expressions for the mean first-encounter times of Markovian walkers under these dynamics. Figure \ref{Fig_1} illustrates the dynamics for two agents navigating a network (for instance a street network) and moving randomly between adjacent nodes. At each time step, the walkers reset simultaneously to their respective home nodes $r_1$ and $r_2$ with probability $\gamma$. With the complementary probability $1 - \gamma$, they follow their random-walk dynamics governed by the transition matrices $\mathbf{W}^{(1)}$ and $\mathbf{W}^{(2)}$. Although the walkers do not interact directly, a central question is whether and when they \emph{encounter} one another; that is, when both walkers occupy the same node at the same time. Our main objective is to analytically determine how simultaneous resetting influences the mean number of time steps required for the walkers to coincide at a given node $j$ for the first time.
\\[2mm]
The paper is organized as follows. In Sec. \ref{Sec_GeneralTheory} we present the general theoretical framework, including the mathematical formalism employed to study the dynamics of $S$ random walkers with simultaneous resetting on a network, and derive the master equation governing the evolution of the occupation probability in the configuration space. We then obtain general expressions for the mean first-encounter times and reformulate them in terms of the spectral properties of the underlying dynamics without resetting, highlighting the role of the eigenvalues and eigenvectors of the transition matrices $\mathbf{W}^{(s)}$ with $s=1,\ldots,S$. We also establish general criteria that determine when resetting becomes advantageous and therefore an optimal, nonzero resetting probability exists. In Sec. \ref{Sec_Results} we present numerical results for representative regular and heterogeneous network topologies, encounters involving two or more walkers, and scenarios combining local random walks with nonlocal Lévy-flight dynamics, thereby illustrating the breadth and robustness of the theory. As an additional test, we analyze two normal random walkers under simultaneous and independent resetting protocols, elucidating how correlations between resetting events influence the optimal search efficiency across different network topologies. Section \ref{Sec_Conclusions} summarizes the main findings, while technical details are exposed in the Appendices.
\section{General theory}
\label{Sec_GeneralTheory}
In this section, we present a general theoretical framework for the collective dynamics of multiple non-interacting random walkers on networks under simultaneous stochastic resetting. We derive the master equation governing the time evolution of the system, which provides the basis for the analytical calculation of encounter and first-passage properties. By exploiting the spectral properties of the transition matrix with resetting, we obtain explicit expressions for the stationary distribution and the mean first-encounter times, and establish general conditions under which resetting enhances encounter processes.
\subsection{Master equation}
\label{Sec_Simul_RW_reset}
We study the dynamics of $S$ random walkers on a connected network with $N$ nodes, denoted by the set $\mathcal{V}=\{1,2,\ldots,N\}$. The walker $s$ evolves according to a transition matrix $\mathbf{W}^{(s)}$ of size $N \times N$, whose elements $(\mathbf{W}^{(s)})_{ij}=w^{(s)}_{i\to j}$ specify the probability of moving from node $i$ to node $j$. A matrix $\hat{\mathcal{W}}$ that captures the collective dynamics of the $S$ non-interacting random walkers under synchronous motion is defined as \cite{riascos_multiple2020}
\begin{equation}\label{Wmatrix_S}
	\hat{\mathcal{W}} \equiv \bigotimes_{s=1}^S \mathbf{W}^{(s)}
	= \mathbf{W}^{(1)} \otimes \mathbf{W}^{(2)} \otimes \cdots \otimes \mathbf{W}^{(S)},
\end{equation}
where $\otimes$ denotes the tensor product of matrices. For convenience, we introduce the notation $\vec{i} \equiv (i_1, i_2, \ldots, i_S) \in \mathcal{V}^S$, where $i_1, i_2, \ldots, i_S \in \{1,2,\ldots,N\}$ denote the positions of the walkers, i.e. $i_s$ indicates the node occupied by walker $s$ on the network. Using this notation the elements of $\hat{\mathcal{W}}$ in Eq. (\ref{Wmatrix_S}) are $\hat{\mathcal{W}}_{\vec{i} \to \vec{j}}\equiv \prod_{s=1}^S w^{(s)}_{i_s\to j_s}$ and give the probability to pass from configuration $\vec{i}$ to $\vec{j}$. In addition
\begin{equation}
	\sum_{\vec{l}\in \mathcal{V}^S} \hat{\mathcal{W}}_{\vec{i} \to \vec{l}}
	= \prod_{s=1}^S \sum_{l_s=1}^N w^{(s)}_{i_s \to l_s}
	= \prod_{s=1}^S 1 = 1.
\end{equation}
At each discrete time step $t=1,2,\ldots$, all walkers update their positions independently and simultaneously according to one of two possible actions: with probability $1-\gamma$, they all perform random jumps on the network, while with probability $\gamma$, they reset simultaneously to the fixed nodes $\vec{r}=(r_1,r_2,\ldots,r_S)$. The probability $\mathcal{P}(\vec{i},\vec{j},\gamma;t)$ of finding the $S$ walkers at nodes $\vec{j}$ at time $t$, given their initial positions $\vec{i}$ at $t=0$, satisfies the master equation
\begin{equation}
	\mathcal{P}(\vec{i} ,\vec{j},\gamma;t+1)=(1-\gamma)\sum_{\vec{l}\in \mathcal{V}^S} \mathcal{P}(\vec{i} ,\vec{l},\gamma;t)\hat{\mathcal{W}}_{\vec{l} \to \vec{j}}+\gamma \delta_{\vec{r},\vec{j}},
	\label{masterRW_W_reset}
\end{equation}
where $\delta_{\vec{r},\vec{j}} = \delta_{r_1, j_1}\delta_{r_2, j_2}\cdots\delta_{r_S, j_S}$, with $\delta_{r,j}$ denoting the Kronecker delta. The first term on the right-hand side of Eq.~(\ref{masterRW_W_reset}) accounts for the collective hops governed by the transition probabilities $\hat{\mathcal{W}}$, while the second term represents the resetting process to $\vec{r}$. Alternatively, the dynamics under resetting can be described by the transition probability matrix $\hat{\mathbf{\Pi}}(\vec{r};\gamma)$, defined as
\begin{equation}\label{Def_Pi_matrix}
	\hat{\mathbf{\Pi}}(\vec{r};\gamma)\equiv (1-\gamma)\hat{\mathcal{W}}+\gamma\hat{\mathbf{\Theta}}(\vec{r}).
\end{equation}
Here, $\hat{\mathbf{\Theta}}(\vec{r})$ describes resetting to the nodes $\vec{r}=(r_1,r_2,\ldots,r_S)$ and is given by 
\begin{equation}
	\hat{\mathbf{\Theta}}(\vec{r}) \equiv \bigotimes_{s=1}^S \mathbf{\Theta}(r_s),
\end{equation}
where the matrix $\mathbf{\Theta}(r_s)$ has elements $\Theta_{lm}(r_s) = \delta_{m,r_s}$. In this way, using the definition of $\hat{\mathbf{\Pi}}(\vec{r};\gamma)$, Eq.~(\ref{masterRW_W_reset}) can be written in the more compact form
\begin{equation}
	\mathcal{P}(\vec{i} ,\vec{j},\gamma;t+1)=\sum_{\vec{l}\in \mathcal{V}^S} \mathcal{P}(\vec{i} ,\vec{l},\gamma;t)\hat{\mathbf{\Pi}}(\vec{r};\gamma)_{\vec{l} \to \vec{j}} \; .
	\label{masterRrw}
\end{equation}
%
\subsection{Mean first encounter times}
The matrix $\hat{\mathbf{\Pi}}(\vec{r};\gamma)$ fully characterizes the collective dynamics with resetting, ensuring that all walkers can visit every node of the network as long as the resetting probability satisfies $0 \leq \gamma < 1$. The matrix $\mathbf{\Pi}(\vec{r};\gamma)$ is stochastic, i.e.,
\begin{align}\nonumber
\sum_{\vec{l}\in \mathcal{V}^S} \hat{\mathbf{\Pi}}(\vec{r};\gamma)_{\vec{i} \to \vec{l}}
&= (1-\gamma)\sum_{\vec{l}\in \mathcal{V}^S} \hat{\mathcal{W}}_{\vec{i}\to \vec{l}}
+ \gamma \sum_{\vec{l}\in \mathcal{V}^S} \hat{\mathbf{\Theta}}(\vec{r})_{\vec{i}\to \vec{l}}\\
&= (1-\gamma) + \gamma = 1,
\end{align}
where we use the result 
\begin{equation}
\sum_{\vec{l}\in \mathcal{V}^S} \hat{\mathbf{\Theta}}(\vec{r})_{\vec{i}\to \vec{l}}= \prod_{s=1}^S \sum_{l_s=1}^N \delta_{r_s,l_s}= \prod_{s=1}^S 1 = 1.
\end{equation}
In addition, the eigenvalues and eigenvectors of $\hat{\mathbf{\Pi}}(\vec{r};\gamma)$ provide key insights into the system’s dynamics. In particular, knowledge of these spectral properties allows one to calculate the occupation probability $\mathcal{P}(\vec{i},\vec{j},\gamma;t)$, including the stationary distribution at $t \to \infty$, as well as the mean first-passage time to any configuration $\vec{j}$, starting from configuration $\vec{i}$. 
\\[2mm]
On the other hand, Eq. (\ref{masterRrw}) has a direct analogy with the dynamics of a single walker \cite{reviewjcn_2021}. This similarity allows for an analytical treatment of the problem of $S$ non-interacting random walkers, particularly for calculating the mean number of steps required to first reach a given configuration. To this end, we begin by introducing a compact notation for the eigenvalues and eigenvectors of $\hat{\mathbf{\Pi}}(\vec{r};\gamma)$, which play a central role in the analysis of the master equation (\ref{masterRrw}). In the following, we use Dirac's notation for the eigenvectors, writing
\begin{equation}\label{Weigenvector0}
	\hat{\mathbf{\Pi}}(\vec{r};\gamma)|\psi_{\vec{i}}\rangle=\zeta_{\vec{i}}|\psi_{\vec{i}}\rangle \qquad \vec{i}\in \mathcal{V}^S,
\end{equation}
where $\zeta_{\vec{i}}$ denotes the eigenvalues of the transition operator $\hat{\mathbf{\Pi}}(\vec{r};\gamma)$ with the corresponding set of right eigenvectors $\{ |\psi_{\vec{i}}\rangle \}_{\vec{i}\in \mathcal{V}^S}$. The associated left eigenvectors, denoted as $\{\langle\bar{\psi}_{\vec{i}}|\}_{\vec{i}\in \mathcal{V}^S}$, satisfy
\begin{equation}
	\langle\bar{\psi}_{\vec{i}}|\hat{\mathbf{\Pi}}(\vec{r};\gamma)=
	\zeta_{\vec{i}}\langle\bar{\psi}_{\vec{i}}|.
\end{equation}
The eigenvectors fulfill the orthonormalization property
\begin{equation}
	\langle\bar{\psi}_{\vec{i}}|\psi_{\vec{j}}\rangle
	= \delta_{i_1,j_1}\delta_{i_2,j_2}\cdots\delta_{i_S,j_S}
	\equiv \delta_{\vec{i},\vec{j}},
\end{equation}
and the completeness relation
\begin{equation}
	\sum_{\vec{l}\in \mathcal{V}^S}\left|\psi_{\vec{l}}\right\rangle
	\left\langle\bar{\psi}_{\vec{l}}\,\right|
	= \mathbf{I}^{\otimes S},
\end{equation}
where $\mathbf{I}$ denotes the $N \times N$ identity matrix. The components of the left and right eigenvectors are expressed using the vectors $|\vec{i}\rangle \equiv |i_1,i_2,\ldots,i_S\rangle = |i_1\rangle \otimes |i_2\rangle \otimes \cdots \otimes |i_S\rangle$ and $\langle \vec{i}| = |\vec{i}\rangle^\dagger$, where $\dagger$ denotes the Hermitian conjugate. In this notation, $|i_s\rangle$ denotes the $N$-dimensional vector with all entries equal to 0, except the $i_s^{\rm th}$ one, with entry 1.
\\[2mm]
We can now express the time evolution $\mathcal{P}(\vec{i},\vec{j},\gamma;t)$ of the $S$-walker system in terms of the eigenvalues and the left and right eigenvectors of $\hat{\mathbf{\Pi}}(\vec{r};\gamma)$. By iterating $t$ times Eq.~(\ref{masterRrw}) we have,
\begin{align}\nonumber
\mathcal{P}(\vec{i},\vec{j},\gamma;t)
	&= \langle \vec{i}|\hat{\mathbf{\Pi}}(\vec{r};\gamma)^t|\vec{j}\rangle
	= \sum_{\vec{l}\in \mathcal{V}^S}\langle \vec{i}|\hat{\mathbf{\Pi}}(\vec{r};\gamma)^t|\psi_{\vec{l}}\rangle
	\langle\bar{\psi}_{\vec{l}}|\vec{j}\rangle \\
	&= \sum_{\vec{l}\in \mathcal{V}^S}  \zeta_{\vec{l}}^t \langle \vec{i}\rpsivec{l}\lpsivec{l}\vec{j}\rangle.
	\label{Pijsimul_spect}
\end{align}
The stationary distribution is then given by
\begin{align}\nonumber
	\mathcal{P}^{\infty}_{\vec{j}}(\vec{i},\gamma)
	&\equiv \lim_{T\to \infty}\frac{1}{T}\sum_{t=0}^{T} \mathcal{P}(\vec{i},\vec{j},\gamma;t)\\
	&= \sum_{\vec{l}\in \mathcal{V}^S} \delta_{\zeta_{\vec{l}},1}\,
	\langle \vec{i}\rpsivec{l}\lpsivec{l}\vec{j}\rangle ,
	\label{Pinf_multiple}
\end{align}
where only the largest eigenvalue $\zeta=1$ contributes in the long-time limit. It is therefore essential to characterize the degeneracy of $\zeta=1$ given by
\begin{equation}
	\kappa \equiv \sum_{\vec{l}\in \mathcal{V}^S} \delta_{\zeta_{\vec{l}},1}.
\end{equation}
A value $\kappa=1$ indicates that all initial configurations can reach any final state in a finite time, meaning that the system is irreducible. Conversely, $\kappa>1$ implies the existence of initial conditions that cannot reach certain final states, resulting in a stationary distribution with zero probability for those states \cite{riascos_multiple2020}. We define the set $\mathcal{D} \equiv \{\vec{l}\in\mathcal{V}^S : \zeta_{\vec{l}}=1\}$ and its complement $\mathcal{D}^\mathrm{c} \equiv \mathcal{V}^S \setminus \mathcal{D}$. The stationary distribution in Eq.~(\ref{Pinf_multiple}) can thus be expressed as
\begin{equation}\label{P_infty_vec_general}
	\mathcal{P}^{\infty}_{\vec{j}}(\vec{i},\gamma)
	= \sum_{\vec{l}\in \mathcal{D}} \langle \vec{i}\rpsivec{l}\lpsivec{l}\vec{j}\rangle .
\end{equation}
With this formalism, we can compute the average time (where the dependence on $\gamma$ is made explicit),
\begin{equation}
\langle T(\vec{i},\vec{j},\gamma)\rangle \equiv \langle T(i_1,i_2,\ldots,i_S,j_1,j_2,\ldots,j_S,\gamma)\rangle
\end{equation}
required for the walkers to simultaneously reach, for the first time, the nodes described by $\vec{j}$ when starting from $\vec{i}$ at $t=0$. A detailed derivation can be found in Ref. \cite{riascos_multiple2020} and is recalled in Appendix \ref{Appendix_FPT} for completeness, leading to the expression, for $\vec{i}\neq \vec{j}$
\begin{equation}\label{TijSpect}
	\langle T(\vec{i},\vec{j},\gamma)\rangle
	= \frac{1}{\mathcal{P}_{\vec{j}}^\infty(\vec{i},\gamma)}
	\sum_{\vec{l}\in \mathcal{D}^\mathrm{c}}
	\frac{\langle \vec{j} |\psi_{\vec{l}}\rangle \langle\bar{\psi}_{\vec{l}}|\vec{j}\rangle -
		\langle \vec{i} |\psi_{\vec{l}}\rangle \langle\bar{\psi}_{\vec{l}}|\vec{j}\rangle}{1-\zeta_{\vec{l}}}\,.
\end{equation}
In Eqs. (\ref{P_infty_vec_general}) and (\ref{TijSpect}), the dependence on $\vec{r}$ of the eigenvectors and stationary distribution is implicit. The special case $\vec{i}= \vec{j}$ gives the mean first-return time to the initial condition $\langle T(\vec{i},\vec{i},\gamma)\rangle=1/\mathcal{P}_{\vec{i}}^\infty(\vec{i},\gamma)$, which is Kac's lemma on the mean recurrence time of discrete processes \cite{kac1947on}.
\\[2mm]
In particular, if $j_1, j_2, \ldots, j_S = j$, $\langle T(\vec{i},\vec{j},\gamma)\rangle$ gives the mean first-encounter time (MFET) at node $j$, that is, the average time elapsed until all the walkers coincide for the first time at $j$.
\subsection{MFET in terms of the dynamics without reset}
In this section, we explore the connection between the eigenvalues and eigenvectors of $\hat{\mathbf{\Pi}}(\vec{r};\gamma)$ introduced above and those of the collective dynamics without resetting defined by  $\hat{\mathcal{W}}$. This will allow us to rewrite the result in Eq. (\ref{TijSpect}) in a more explicit way, involving the resetting probability $\gamma$.
\\[2mm]
We assume that each individual transition matrix $\mathbf{W}^{(s)}$ is diagonalizable. Under this assumption, we introduce the $l^{\rm th}$  eigenvector of the walker $s$
\begin{equation}\label{Weigenvector}
	\mathbf{W}^{(s)}|\phi_l^{(s)}\rangle = \lambda_l^{(s)}|\phi_l^{(s)}\rangle, 
	\qquad s = 1, 2, \ldots, S,
\end{equation}
where $\{\lambda_l^{(s)}\}_{l=1}^{N}$ denote the eigenvalues of the transition matrix $\mathbf{W}^{(s)}$, with the corresponding set of right eigenvectors $\{|\phi_l^{(s)}\rangle\}_{l=1}^{N}$ \cite{reviewjcn_2021}. It is convenient to consider any combination of the individual eigenstates through the vector $\vec{l}$ of components $(l_1,\ldots,l_S)$ with 
 $1\le l_s\le N$, and by  defining 
\begin{equation}
	|\phi_{\vec{l}}\rangle \equiv |\phi_{l_1}^{(1)}\rangle \otimes |\phi_{l_2}^{(2)}\rangle \otimes \cdots \otimes |\phi_{l_S}^{(S)}\rangle 
	= \bigotimes_{s=1}^S |\phi_{l_s}^{(s)}\rangle\, .
\end{equation}
From Eqs.~(\ref{Wmatrix_S}) and (\ref{Weigenvector}),
\begin{equation}\label{SpectM}
	\hat{\mathcal{W}}|\phi_{\vec{l}}\rangle = \lambda_{\vec{l}}|\phi_{\vec{l}}\rangle,
\end{equation}
where, using the definition in Eq.~(\ref{Wmatrix_S}) for synchronous random walkers, the eigenvalues of $\hat{\mathcal{W}}$ are obtained as
\begin{equation}\label{zeta_S}
	\lambda_{\vec{l}} \equiv \prod_{s=1}^S \lambda_{l_s}^{(s)}.
\end{equation}
In addition, since the eigenvalues of $\mathbf{W}^{(s)}$ satisfy $|\lambda_{l_s}^{(s)}|\leq 1$, it follows that $|\zeta_{\vec{l}}|\leq 1$ for all ${\vec{l}}\in \mathcal{V}^S$. To proceed further, we also need the set of left eigenvectors. For the individual transition matrix $\mathbf{W}^{(s)}$, we have 
$\langle \bar{\phi}^{(s)}_l|\mathbf{W}^{(s)} = \lambda_l^{(s)}\langle \bar{\phi}^{(s)}_l|$.
Then, defining $\langle\bar{\phi}_{\vec{l}}| \equiv \bigotimes_{s=1}^S \langle\bar{\phi}_{l_s}^{(s)}|$, we obtain
\begin{equation}
	\langle\bar{\phi}_{\vec{l}}|\hat{\mathcal{W}} = \zeta_{\vec{l}}\langle\bar{\phi}_{\vec{l}}|.
\end{equation}
Since each set of eigenvectors of $\mathbf{W}^{(s)}$ satisfies 
$\delta_{m,n} = \langle\bar{\phi}^{(s)}_m|\phi^{(s)}_n\rangle$ and 
$\mathbf{I} = \sum_{l=1}^N |\phi^{(s)}_l\rangle \langle \bar{\phi}^{(s)}_l |$ \cite{reviewjcn_2021}, 
it follows from the definitions of $|\phi_{\vec{l}}\rangle$ and $\langle\bar{\phi}_{\vec{l}'}|$ for arbitrary $\vec{l}$ and $\vec{l}'$ that the eigenvectors of $\hat{\mathcal{W}}$ satisfy the orthonormalization condition 
$\langle\bar{\phi}_{\vec{l}}|\phi_{\vec{l}'}\rangle = \delta_{\vec{l},\vec{l}'}$ 
and the completeness relation 
$\sum_{\vec{l}\in \mathcal{V}^S} |\phi_{\vec{l}}\rangle \langle\bar{\phi}_{\vec{l}}| = \mathbf{I}^{\otimes S}$.
\\[2mm]
In the following, we denote the largest eigenvalue of $\mathbf{W}^{(s)}$ as $\lambda_1^{(s)} = 1$. 
According to the Perron--Frobenius theorem, this eigenvalue is unique, and its associated eigenvector determines the stationary distribution of each walker through the relation 
$P^{s,\infty}_{j} = \langle i|\phi_1^{(s)}\rangle \langle\bar{\phi}^{(s)}_1|j\rangle$. 
Notably, this distribution is independent of the initial node $i$, since $\langle i|\phi_1^{(s)}\rangle$ remains constant \cite{reviewjcn_2021}.
\\[2mm]
We can now express the eigenvalues and eigenvectors of $\hat{\mathbf{\Pi}}(\vec{r};\gamma)$ in terms of the eigenvalues and the left and right eigenvectors of $\hat{\mathcal{W}}$. For brevity, we only present the main results here; the interested reader can find a detailed derivation in Appendix \ref{Appendix_Spec}, which is very similar to the single walker case analyzed in \cite{ResetNetworks_PRE2020}. The eigenvalues of  $\hat{\mathbf{\Pi}}(\vec{r};\gamma)$  are given by
\begin{equation}\label{eigvals_zeta}
	\zeta_{\vec{l}} = \left\{
	\begin{array}{ll}
		1 \qquad & \mathrm{if}\qquad \vec{l}=\vec{1} \\
		(1-\gamma)\lambda_{\vec{l}} \qquad &  \mathrm{if}\qquad \vec{l}\in \mathcal{I}, \\
	\end{array}
	\right.
\end{equation}
where $\vec{1}\equiv (1,1,\ldots, 1)$ and  $\mathcal{I} = \mathcal{V}^S \setminus \{\vec{1}\}$. The result in Eq. (\ref{eigvals_zeta})  reveals that the eigenvalues are independent of the choice of the resetting nodes $\vec{r}$. 
\\[2mm]
On the other hand, the right eigenvectors of  $\mathbf{\Pi}(r;\gamma)$ are given by $	\rpsivec{1}=\rphivec{1}$, whereas for $\vec{l}\in \mathcal{I}$,
\begin{equation}\label{Eigen_right_reset}
	\rpsivec{l}=\rphivec{l}-\frac{\gamma}{1-(1-\gamma)\lambda_{\vec{l}}}\frac{\langle \vec{r}\rphivec{l}}{\langle \vec{r}\rphivec{1}} \rphivec{1} .
\end{equation}
Similarly, the left eigenvectors are given by $	\lpsivec{l}=\lphivec{l}$ for $\vec{l}\in \mathcal{I}$, whereas for $ \vec{l}= \vec{1}$, 
\begin{equation}\label{Eigen_left_reset}
	\lpsivec{1}=\lphivec{1}
	+\sum\limits_{\vec{m}\in  \mathcal{I} }\frac{\gamma}{1-(1-\gamma)\lambda_{\vec{m}}}\frac{\langle \vec{r}\rphivec{m}}{\langle \vec{r}\rphivec{1}} \lphivec{m}.
\end{equation}
In this notation, we deduce the occupation probability of the process from Eq. (\ref{Pijsimul_spect}), 
\begin{multline}
\mathcal{P}(\vec{i},\vec{j},\gamma;t)=\mathcal{P}^{\infty}_{\vec{j}}(\vec{r},\gamma)+\sum_{\vec{l}\in  \mathcal{I} }(1-\gamma)^t\lambda_{\vec{l}}^t\\
\times\left[\langle \vec{i}\rphivec{l}\lphivec{l}\vec{j}\,\rangle-\gamma\frac{\langle\vec{r}\rphivec{l}\lphivec{l}\vec{j}\,\rangle}{1-(1-\gamma)\lambda_{\vec{l}}} \right]. \label{Pijspect}
\end{multline}
The first term on the right-hand side of Eq.~(\ref{Pijspect}) is the long-time stationary distribution 
$\mathcal{P}^{\infty}_{\vec{j}}(\vec{r},\gamma) = \langle\vec{i}\rpsivec{1}\lpsivec{1}\vec{j}\,\rangle$, 
since, according to Eq.~(\ref{eigvals_zeta}), for $0 < \gamma < 1$ the largest eigenvalue $\zeta_{\vec{1}} = 1$ is unique. Here, we have explicitly indicated the dependence of the stationary distribution on $\vec{r}$.
Therefore, using Eq.~(\ref{Eigen_left_reset}) for $\lpsivec{1}$ and the identity $\rpsivec{1} = \rphivec{1}$ presented above, we obtain
\begin{equation}\label{Pinfvectors_1}
	\mathcal{P}^{\infty}_{\vec{j}}(\vec{r},\gamma)=\mathcal{P}^{\infty}_{\vec{j}}(0)+\gamma\sum_{\vec{l}\in  \mathcal{I}}\frac{\langle\vec{r}\rphivec{l}\lphivec{l}\vec{j}\,\rangle}{1-(1-\gamma)\lambda_{\vec{l}}}.
\end{equation}
In this expression, $\mathcal{P}^{\infty}_{\vec{j}}(0) \equiv  \langle\vec{i}\rphivec{1}\lphivec{1}\vec{j}\,\rangle$ corresponds to the stationary distribution of the multiple random walks without resetting when the eigenvalue $\lambda_{\vec{1}} = 1$ is non-degenerate.
\\[2mm]
Furthermore, applying the same procedure to Eq.~(\ref{TijSpect}) for the MFET $\langle T(\vec{i},\vec{j},\gamma)\rangle$ of random walkers starting from the nodes $\vec{i}$ and reaching simultaneously the target node $\vec{j} = (j,j,\ldots,j)$ for the first time, while being subject to stochastic resetting to the nodes $\vec{r}$, we obtain for $\vec{i} \neq \vec{j}$
\begin{equation}\label{MFPT_resetSM}
	\langle T(\vec{i},\vec{j}\,,\gamma)\rangle
=\frac{1}{\mathcal{P}^{\infty}_{\vec{j}}(\vec{r},\gamma)}
\sum_{\vec{l}\in \mathcal{I}}\frac{\langle \vec{j} |\phi_{\vec{l}}\rangle\langle\bar{\phi}_{\vec{l}}\,|\vec{j}\rangle-
	\langle \vec{i} |\phi_{\vec{l}}\rangle\langle\bar{\phi}_{\vec{l}}\,|\vec{j}\rangle}{1-(1-\gamma)\lambda_{\vec{l}}}
\end{equation}
valid for $0 < \gamma < 1$. In the case $\gamma = 0$, Eq.~(\ref{MFPT_resetSM}) with $\mathcal{P}^{\infty}_{\vec{j}}(\vec{r},0) = \sum_{\vec{l}\in \mathcal{D}} \langle \vec{i}\rphivec{l}\lphivec{l}\vec{j}\rangle$ becomes the MFET obtained for the dynamics without resetting, as studied in Ref.~\cite{riascos_multiple2020}. This modification in the stationary distribution accounts for the degeneracy of the largest eigenvalue of $\hat{\mathcal{W}}$ (see Ref. \cite{riascos_multiple2020} for details).
\subsection{Conditions under which resetting becomes advantageous}

In the previous sections, we considered multiple discrete-time random walkers with simultaneous resetting to the nodes $\vec{r}$ and analytically explored the effect of the resetting probability. In this section, we establish a general condition that determines when resetting is advantageous, i.e., reduces the MFET compared to the case $\gamma=0$. This issue has been previously discussed for single random walks or processes \cite{reuveni2016optimal,pal2017first,pal2022inspection,pal2024random,bonomo2021first,riascos_jpa_2022}.
The detailed derivation is presented for multiple walkers in Appendix \ref{Appendix_Adv}; here, we summarize the main ideas and results. The whole approach is valid in cases when the largest eigenvalue of $\hat{\mathcal{W}}$ is non-degenerate.
\\[2mm]
We start by considering the MFET for the dynamics with resetting to the initial configuration in Eq.~(\ref{MFPT_resetSM}), i.e., we set $\vec{r}=\vec{i}$. The MFET can be expressed in this case more compactly as
\begin{equation}
	\langle T(\vec{i},\vec{j}\,,\gamma)\rangle=\frac{\mathcal C_{\vec i,\vec j}(\gamma)}
	{P_{\vec{j}}^\infty(0)+\gamma\,\mathcal S_{\vec i,\vec j}(\gamma)},
\end{equation}
where $\mathcal C_{\vec i,\vec j}(\gamma)$ and $\mathcal S_{\vec i,\vec j}(\gamma)$ denote spectral sums depending on the eigenvalues and eigenvectors of the transition matrices without resetting, and are defined in Appendix \ref{Appendix_Adv}. The optimal resetting probability $\gamma^\star$ follows from the condition
\begin{equation}
	\frac{d}{d\gamma}\langle T(\vec{i},\vec{j}\,,\gamma)\rangle\Big|_{\gamma=\gamma^\star}=0,
\end{equation}
which can be solved numerically once the spectral data of the transition matrices are known.
\\[2mm]
A notable simplification arises when the above condition is expressed in terms of the first two moments of the first-encounter times. Using the relations between the spectral coefficients and the moment-generating structure of the process, for $\vec{i}\neq \vec{j}$ one obtains the identity 
\begin{equation}
	\langle T^2(\vec{i},\vec{j}\,,\gamma^\star)\rangle
	= \langle T(\vec{i},\vec{j}\,,\gamma^\star)\rangle
	+ 2\,\langle T(\vec{i},\vec{j}\,,\gamma^\star)\rangle^2.
	\label{eq:T2_opt_simplified}
\end{equation}
Introducing the coefficient of variation of the first encounter time, defined as
\begin{equation}\label{z_definition}
	z(\vec{i},\vec{j}\,,\gamma)\equiv 
	\frac{\sqrt{\langle T^2(\vec{i},\vec{j}\,,\gamma)\rangle - \langle T(\vec{i},\vec{j}\,,\gamma)\rangle^2}}
	{\langle T(\vec{i},\vec{j}\,,\gamma)\rangle},
\end{equation}
Eq.~(\ref{eq:T2_opt_simplified}) yields a general criterion for the fluctuations of the encounter time at optimality
\begin{equation}
z^2(\vec{i},\vec{j}\,,\gamma^\star)=1+\frac{1}{\langle T(\vec{i},\vec{j}\,,\gamma^\star)\rangle},
	\qquad \vec{i}\neq \vec{j}.
	\label{eq:z_opt_condition}
\end{equation}
A second criterion determines whether the introduction of resetting is advantageous at all. Clearly, a small amount of resetting ($\gamma$ small) accelerates the encounters if
\begin{equation}\label{dTneg}
	\frac{d}{d\gamma}\langle T(\vec{i},\vec{j}\,,\gamma)\rangle\Big|_{\gamma\to 0}<0,
\end{equation}
a condition that determines whether the introduction of an infinitesimal resetting probability decreases the MFET. Therefore, Eq. (\ref{dTneg}) provides a \emph{sufficient local criterion} for the existence of a nonzero optimal resetting probability. However, this criterion is not necessary, as a non-trivial globally optimal resetting rate may exist even when the initial slope is positive (see, e.g., Ref.~\cite{Pal_PRR_2016}). Indeed, since the MFET diverges in the limit $\gamma\rightarrow1$, a global minimum must exist at some $0<\gamma^\star<1$. This criterion is independent of the specific dynamics between resetting events and depends only on the first two moments of the first-encounter time in the absence of resetting. Consequently, it applies more broadly to any random-walk dynamics whose underlying evolution defines an ergodic discrete-time Markov process with a non-degenerate stationary state.
\\[2mm]
Expressed in terms of the moments of the first-encounter times of the underlying process (with $\gamma=0$), Eq. (\ref{dTneg}) becomes
\begin{equation}
z^2(\vec{i},\vec{j}\,,0)>1+\frac{1}{\langle T(\vec{i},\vec{j}\,,0)\rangle},
\qquad \vec{i}\neq \vec{j}.
	\label{eq:z_smallgamma_condition}
\end{equation}
Using the definition of $z(\vec{i},\vec{j},\gamma)$ introduced in Eq. (\ref{z_definition}), let us define
\begin{equation}
	\Lambda\equiv\frac{\braket{T^2(\vec{i},\vec{j}\,,0)}-\braket{T(\vec{i},\vec{j}\,,0)}^2}{\braket{T(\vec{i},\vec{j}\,,0)}^2}-\frac{1}{\braket{T(\vec{i},\vec{j}\,,0)}}-1\,.
	\label{Lambda_def}
\end{equation}
As a consequence of Eq. (\ref{eq:z_smallgamma_condition}), the value taken by $\Lambda$ in Eq. (\ref{Lambda_def}) for a given system determines whether the introduction of resetting is advantageous. When $\Lambda>0$, an infinitesimal resetting probability reduces the MFET, implying the existence of a nonzero optimal resetting probability $\gamma^\star$. In contrast, when $\Lambda<0$, a slight introduction of resetting increases the MFET, and, in principle, the optimal resetting probability is $\gamma^{\star}=0$. In this manner, Eqs.~(\ref{eq:z_opt_condition})-(\ref{Lambda_def}) provide general criteria characterizing optimal resetting and the beneficial introduction of simultaneous reset events in discrete-time ergodic Markov processes. These conditions hold for arbitrary network topologies and transition rules, and therefore offer a unified framework to assess the impact of resetting on first-encounter times. The calculation of the second moment appearing in Eq. (\ref{Lambda_def}) from the spectral properties of $\hat{\mathcal W}$ is exposed in Appendix \ref{Appendix_FPT}.
\section{Results}
\label{Sec_Results}
\begin{figure*}[t!]
	\centering
	\includegraphics[width=0.88\textwidth]{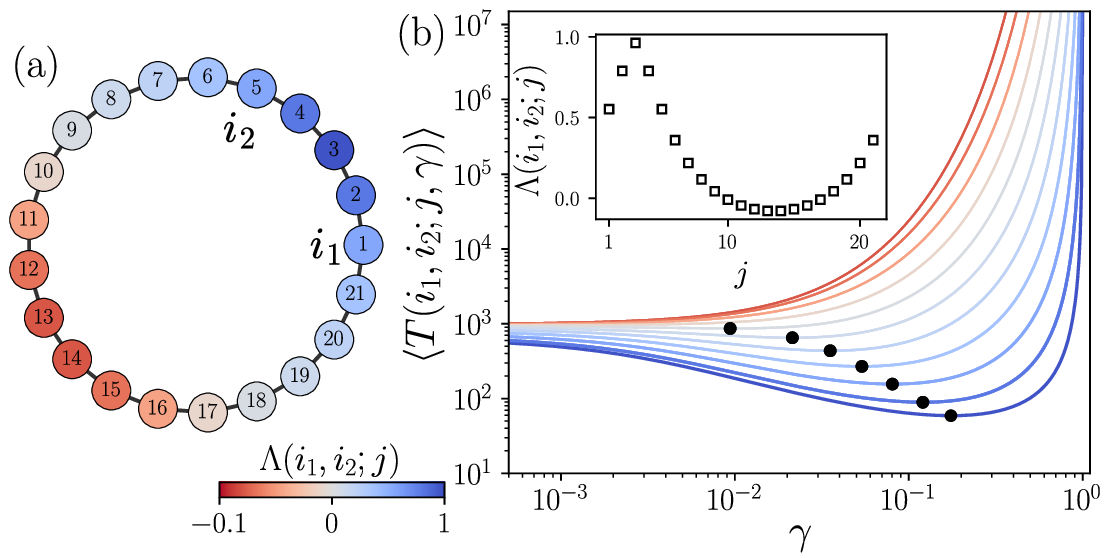}
	\vspace{-4mm}
	\caption{Mean first-encounter times of two normal random walkers with simultaneous resetting on a ring with $N=21$ nodes. (a) Ring topology indicating the initial nodes $i_1$ and $i_2$. (b) The quantity $\langle T(i_1,i_2;j,\gamma)\rangle$ as a function of the simultaneous resetting probability $\gamma$, with the minimal values represented by dots. The inset displays the values of $\Lambda(i_1,i_2;j)$ in Eq. (\ref{Lambda_def}) as a function of $j$. The value of $\Lambda(i_1,i_2;j)$ for each $j$ is also encoded in the color bar, which identifies the nodes in panel (a) and the corresponding curves in panel (b). Further details are provided in the main text.
	}
	\label{Fig_2}
\end{figure*}
In this section, we present numerical results illustrating the impact of simultaneous resetting on the mean first-encounter time of multiple random walkers. We focus on representative network topologies and initial conditions to elucidate the role of the resetting probability and the choice of the target node. The results confirm the theoretical predictions derived in the previous sections and we identify the conditions under which resetting significantly reduces the encounter times.
\\[2mm]
For the mean first-encounter times $\langle T(\vec{i},\vec{j},\gamma)\rangle$ in Eq. (\ref{MFPT_resetSM}), we consider $\vec{j}=(j,j,\ldots,j)$ to describe encounters at node $j$ under simultaneous resetting to the origin. In the following, we use the notation for the MFET $\langle T(\vec{i};j,\gamma)\rangle$ for $S$ random walkers. In the special case $S=2$, we adopt the notation $\langle T(i_1,i_2;j,\gamma)\rangle$ to explicitly emphasize the dependence on the initial conditions $i_1$ and $i_2$.
\subsection{Two random walkers on a ring }
\begin{figure*}[t!]
	\centering
	\includegraphics[width=0.9\linewidth]{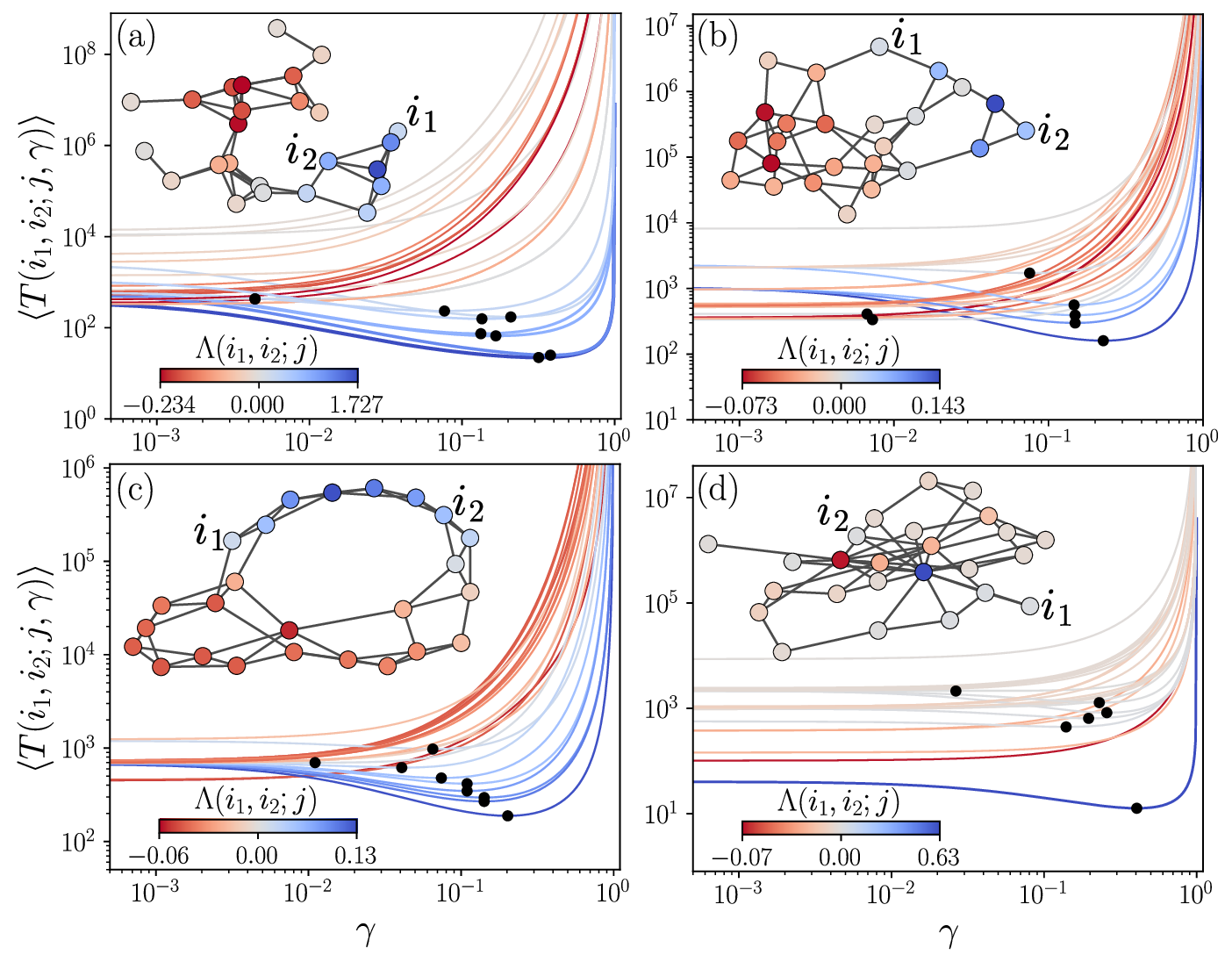}
	\\[-4mm]
	\caption{Mean first-encounter times $\langle T(i_1,i_2;j,\gamma)\rangle$ as a function of the simultaneous resetting probability $\gamma$ for two normal random walkers on heterogeneous networks with $N=25$ nodes. (a) Random geometric graph, (b) Erdős–Rényi network, (c) Watts–Strogatz network, and (d) Barabási–Albert network. The minimal values of $\langle T(i_1,i_2;j,\gamma)\rangle$ are indicated by dots. The curves are color coded according to the values of $\Lambda(i_1,i_2;j)$, as shown in the color bar. The corresponding network and the initial nodes $i_1$ and $i_2$ are displayed in the insets. Further details are provided in the main text.
	}
	\label{Fig_3}
\end{figure*}
We begin by analyzing the dependence of the MFET on the target node $j$ for two normal, unbiased random walkers with identical transition probability matrix and simultaneous resetting to their initial positions on a ring of size $N=21$. For the ring and any other network topology, the normal random walks are defined by the transition probabilities  $w^{(s)}_{i \to j} = A_{ij} / k_i$ for $s=1,2$, where $\mathbf{A}$ is the adjacency matrix ($A_{ij}=1$ if $i$ and $j$ are connected, $0$ otherwise) and $k_i$ is the number of neighbors of node $i$.
\\[2mm]
The results are presented in Fig.~\ref{Fig_2}. Figure~\ref{Fig_2}(a) shows the network topology together with the initial nodes of the random walkers fixed to $i_1=1$ and $i_2=5$.
In Fig.~\ref{Fig_2}(b) we present $\langle T(i_1,i_2;j,\gamma)\rangle$ as a function of $\gamma$, obtained by numerically evaluating Eq.~(\ref{MFPT_resetSM}) for $j=1,2,\ldots,N$. The different nodes $j$ in Fig. \ref{Fig_2}(a) [and their corresponding curves on Fig. \ref{Fig_2}(b)] are color-coded according to the values of $\Lambda$ in Eq. (\ref{Lambda_def}), which determines whether resetting is advantageous or not. In particular, for two random walkers we denote $\Lambda$ as $\Lambda(i_1,i_2;j)$; when $\Lambda(i_1,i_2;j)>0$, the introduction of simultaneous resetting reduces the MFET with respect to its value without resetting, $\langle T(\vec{i};j,\gamma\to 0)\rangle$. Conversely, when $\Lambda(i_1,i_2;j)<0$, a small amount of resetting increases the mean encounter time, and the latter is usually a monotonically increasing function of $\gamma$ all the way to $\gamma=1$. The inset reports the values of the quantity $\Lambda(i_1,i_2;j)$ defined in Eq.~(\ref{Lambda_def}) as a function of $j$, which encapsulates the combined effect of the walkers’ initial positions and the encounter node. The relevance of $\Lambda(i_1,i_2;j)$ lies in the fact that, based on information on the dynamics without resetting only, it allows us to anticipate whether the introduction of resetting can reduce the encounter times. 
As shown by the inset, $\Lambda(i_1,i_2;j)$ exhibits a pronounced dependence on the location of the target node, reflecting the geometric symmetry of the ring and the relative distances between $j$ and the initial nodes $i_1$ and $i_2$. These predictions are confirmed by the behavior of the curves $\langle T(i_1,i_2;j,\gamma)\rangle$ as a function of $\gamma$, for which a well-defined minimum at $\gamma^\star>0$ is observed only in those cases where $\Lambda(i_1,i_2;j)>0$, indicated by dots along the curves.
\\[2mm]
Our findings in Fig.~\ref{Fig_2} reveal that the nodes located closer to the initial positions benefit from resetting. In particular, the node $j=3$ between $i_1$ and $i_2$ exhibits the largest reduction in the mean encounter time, which is consistent with its higher value of $\Lambda(i_1,i_2;j)$. In contrast, the nodes associated with the largest mean encounter times, shown in dark red for $j=13,14$, correspond to situations in which resetting is ineffective. Overall, these results demonstrate that, even in a simple and highly symmetric topology such as the ring, the benefit of simultaneous resetting strongly depends on the choice of the encounter node. The values of $\Lambda(i_1,i_2;j)$ encoded in the color bar thus provide a compact characterization that predicts which targets are more efficiently reached and how the optimal resetting strategy varies across the network.
\begin{figure*}[t!]
	\centering
	\includegraphics[width=0.9\textwidth]{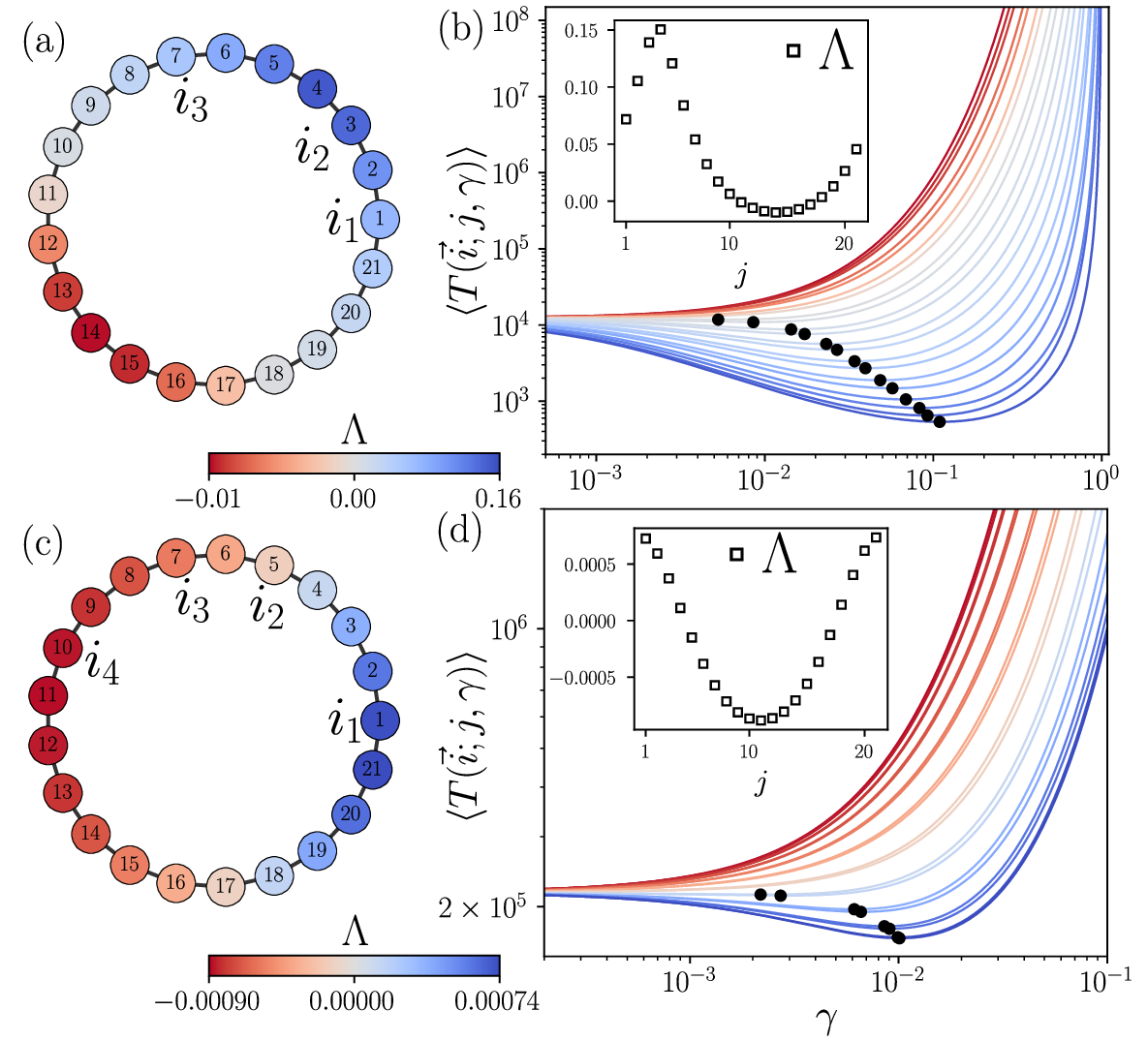}
	\\[-3mm]
	\caption{Mean first-encounter times of three (a,b) and four (c,d) normal random walkers with simultaneous resetting on a ring with $N=21$ nodes. (a) Ring topology showing the initial nodes $i_1$, $i_2$, and $i_3$. (b) Mean first-encounter time $\langle T(\vec{i};j,\gamma)\rangle$, with $\vec{i}=(i_1,i_2,i_3)$, as a function of the simultaneous resetting probability $\gamma$ for every encounter node $j=1,\ldots,N$. Minimal values are indicated by dots. The inset shows the values of $\Lambda$ defined in Eq. ($\ref{Lambda_def}$) as a function of $j$. These values are also encoded in the color bar, which identifies the nodes in panel (a) and the corresponding curves in panel (b). Panels (c) and (d) repeat the analysis for four random walkers with initial conditions $\vec{i}=(i_1,i_2,i_3,i_4)$, as indicated in panel (c).}
	\label{Fig_4}
\end{figure*}
\subsection{Encounter times of two random walkers on heterogeneous structures}
We now extend the analysis beyond regular rings and investigate how network heterogeneity affects the MFET of two random walkers subject to simultaneous resetting. Figure~\ref{Fig_3} summarizes the behavior of $\langle T(i_1,i_2;j,\gamma)\rangle$ on several representative network models with $N=25$ nodes. These topologies include structural disorder and degree heterogeneity, allowing us to assess the robustness of the criterion derived in the previous sections.
\\[2mm]
In Fig.~\ref{Fig_3}(a) we analyze a random geometric graph constructed by placing nodes uniformly at random in two dimensions within the square region $0\leq x,y\leq 1$ and connecting pairs of nodes whose mutual Euclidean distance is smaller than a prescribed threshold \cite{RGG_PRE_Dall_2002}. In this case, we use a connection radius $R=0.22$, which ensures a connected network while preserving strong spatial heterogeneity.  Figure~\ref{Fig_3}(b) shows the results for an Erdős–Rényi random network \cite{ErdosRenyi1959}, where each pair of nodes is connected independently with probability $p=0.09$, yielding a network with average degree $\bar{k}=3.52$. In Fig.~\ref{Fig_3}(c) we consider a Watts–Strogatz network \cite{WattsStrogatz1998} generated from an initial ring with nearest- and next-nearest-neighbor connections and a rewiring probability $p=0.1$ to a randomly chosen node. The introduction of shortcuts breaks the translational invariance of the ring while preserving a high level of local clustering. Finally, Fig.~\ref{Fig_3}(d) corresponds to a scale-free Barabási–Albert network \cite{BarabasiAlbert1999}, constructed using the preferential attachment mechanism with $m=2$, which produces a pronounced degree heterogeneity. The insets in Fig.~\ref{Fig_3} display the specific network realizations and the initial positions $i_1$ and $i_2$ of the walkers used in each case. 
\\[2mm]
Our findings in Fig. \ref{Fig_3} show that, as in the case of the ring, the MFET curves as a function of the resetting probability $\gamma$ display a strong dependence on the target node $j$. For each network realization, different values of $j$ lead to qualitatively distinct behaviors, ranging from a clear reduction of the MFET at an optimal resetting probability $\gamma^\star>0$ to a monotonic increase of $\langle T(i_1,i_2;j,\gamma)\rangle$ with $\gamma$. The minima of the curves, highlighted by dots, signal the existence of an optimal resetting strategy whenever resetting is beneficial.
\\[2mm]
Importantly, this behavior is again consistently captured by the quantity $\Lambda(i_1,i_2;j)$, whose values are encoded in the color bar. As anticipated from the theoretical criterion in Eq.~(\ref{eq:z_smallgamma_condition}), targets with $\Lambda(i_1,i_2;j)>0$ correspond to curves exhibiting a minimum at $\gamma^\star>0$, whereas targets with $\Lambda(i_1,i_2;j)<0$ show no improvement under resetting. This demonstrates that $\Lambda(i_1,i_2;j)$ retains its predictive power even in the presence of strong topological heterogeneity and complex connectivity patterns. Although the detailed shape of the MFET curves depends on the underlying topology, the overall phenomenology remains unchanged across all network classes considered. These results confirm that the spectral-based criterion encoded in $\Lambda(i_1,i_2;j)$ provides a unified framework to anticipate the effect of simultaneous resetting on encounter times, independently of whether the underlying structure is regular or heterogeneous.
\subsection{Encounter times for three and four random walkers}
We now extend the analysis to the case of more than two walkers and investigate how simultaneous resetting affects the MFET when three and four normal random walkers evolve on a ring. Figure~\ref{Fig_4} summarizes the results for a ring of size $N=21$, allowing for a direct comparison with the two-walker case discussed in Fig. \ref{Fig_2}.
\\[2mm]
For three random walkers, Fig.~\ref{Fig_4}(a) displays the ring topology together with the initial positions $\vec{i}=(i_1,i_2,i_3)$. In Fig.~\ref{Fig_4}(b), we show the MFET $\langle T(\vec{i};j,\gamma)\rangle$ as a function of the simultaneous resetting probability $\gamma$ for all possible encounter nodes $j=1,\ldots,N$. As in the two-walker case, the curves exhibit a strong dependence on the location of the encounter node. For a subset of targets, the MFET displays a well-defined minimum at a finite resetting probability, indicating that resetting accelerates the encounter process. These optimal values are highlighted by dots along the curves. The inset in Fig.~\ref{Fig_4}(b) reports the values of the quantity $\Lambda$, defined in Eq.~(\ref{Lambda_def}), as a function of the encounter node $j$. As previously, the sign of $\Lambda$ predicts whether resetting is beneficial. Targets with $\Lambda>0$ correspond to curves that exhibit a minimum at $\gamma>0$, whereas targets with $\Lambda<0$ show a monotonic increase of the MFET with $\gamma$. The same values of $\Lambda$ are encoded in the color bar, which consistently identifies the encounter nodes on the ring and the corresponding MFET curves, as well as their positions in Fig.~\ref{Fig_4}(a).
\\[2mm]
Figures~\ref{Fig_4}(c)-(d) extend this analysis to four random walkers with initial conditions $\vec{i}=(i_1,i_2,i_3,i_4)$. Despite the increased dimensionality of the configuration space, the qualitative behavior remains unchanged. The MFET retains a marked dependence on the encounter node, and the presence or absence of an optimal resetting probability is again predicted by the sign of $\Lambda$.
\\[2mm]
Overall, these results show that the phenomenology observed for two random walkers naturally generalizes to higher-order encounters. Even in the presence of multiple walkers, the quantity $\Lambda$ provides a compact and predictive measure to determine when simultaneous resetting reduces encounter times, highlighting the robustness of the theoretical framework developed in Sec. \ref{Sec_GeneralTheory}.
\subsection{One normal random walker and one following a L\'evy flight}
\begin{figure*}[t!]
	\centering
	\includegraphics[width=1.0\linewidth]{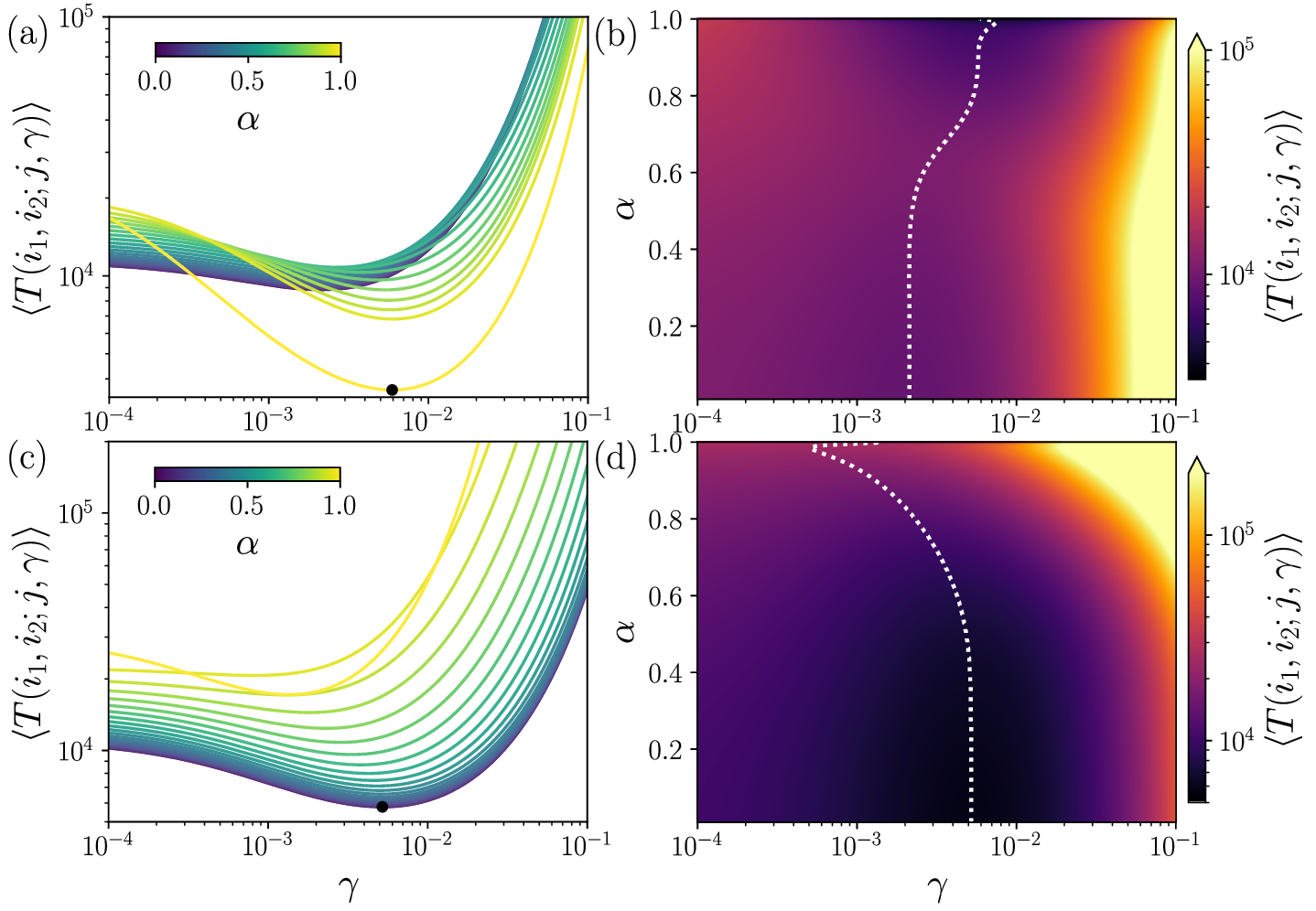}
	\\[-3mm]
	\caption{Encounter times for two random walkers with simultaneous resetting: one following a normal random walk ($s=1$) and the other performing a L\'evy flight ($s=2$). In all cases, the walkers move on a ring of size $N=101$ with initial positions $i_1=1$ and $i_2=21$. (a) MFET $\langle T(i_1,i_2;j,\gamma)\rangle$ as a function of the simultaneous resetting probability $\gamma$ for $j=16$ and different values of $\alpha=0,0.05,\ldots,0.95,1.0$, as indicated by the color bar. The dot marks the minimum encounter time obtained for $\alpha=1$ at $\gamma^\star=0.0059$. (b) MFET as a function of $\gamma$ and $\alpha$ for $j=16$. In this panel, encounter times are encoded in the color bar, and the dashed line indicates the value of $\gamma$ that minimizes $\langle T(i_1,i_2;j,\gamma)\rangle$ for each value of $\alpha$. Panels (c) and (d) repeat the same analysis for $j=92$. In this case, the minimum encounter time is obtained in the limit $\alpha\to0$ at $\gamma^\star=0.0052$. See the main text for details.}
	\label{Fig_5}
\end{figure*}
Previously, we analyzed the encounter times of normal random walkers following the same transition matrix in between resets, where each walker hops from one node to a nearest neighbor chosen with equal probability. However, the theory of  Sec.~\ref{Sec_GeneralTheory} is more general and allows for the analysis of multiple random walkers, each characterized by a different transition matrix. In this section, we explore this more general scenario by considering encounter processes on a ring involving two walkers: a first walker ($s=1$) following the normal local dynamics $w^{(1)}_{i \to j} = A_{ij} / k_i$, and a second walker ($s=2$) performing a nonlocal dynamics known as a L\'evy flight strategy \cite{riascosmateosfd2014,reviewjcn_2021}.
\\[2mm]
L\'evy flights on an arbitrary graph can be generated by considering real powers of the Laplacian matrix $\mathbf{L}$, whose elements are given by $L_{ij}=\delta_{i,j} k_i-A_{ij}$. The transition probabilities for the second random walker are defined as \cite{riascosmateosfd2014}

\begin{equation}\label{wijfrac}
	w_{i\to j}(\alpha)=\delta_{i,j}-\frac{(\mathbf{L}^\alpha)_{ij}}{(\mathbf{L}^\alpha)_{ii}}, \qquad 0<\alpha<1.
\end{equation}
In the limit $\alpha\to1$, this expression recovers the normal random walk with transitions restricted to nearest-neighbor nodes. For $0<\alpha<1$, the transition probabilities in Eq. (\ref{wijfrac}) define a L\'evy flight, characterized by long-range hops in rings with $w_{i\to j}(\alpha)\sim d_{ij}^{-(1+2\alpha)}$ for $d_{ij}\gg1$, where $d_{ij}$ denotes the length of the shortest path between nodes $i$ and $j$. Further details on L\'evy flights and fractional transport on networks can be found in Refs. \cite{riascosmateos2012,riascosmateosfd2014,fractionalbook2019}.
\\[2mm]
Figure~\ref{Fig_5} illustrates the effect of combining two random-walk strategies with simultaneous resetting on a ring of size $N=101$. The normal random walker and the Lévy walker start from initial positions $i_1=1$ and $i_2=21$, respectively. The ring consists of nodes $i=1,2,\ldots,N$, with links connecting nodes $i$ and $i+1$, and periodic boundary conditions are imposed. Panels \ref{Fig_5}(a) and (b) focus on the encounter node $j=16$, located quite close to both resetting points, while panels \ref{Fig_5}(c) and (d) report the results for $j=92$, which is closer to the resetting point of the normal random walker. The L\'evy flight exponent $\alpha$ interpolates between purely nonlocal dynamics in the limit $\alpha\to0$ and the standard nearest-neighbor random walk recovered at $\alpha=1$. 
\\[2mm]
In Fig.~\ref{Fig_5}(a), we show the MFET $\langle T(i_1,i_2; j,\gamma)\rangle$ as a function of the simultaneous resetting probability $\gamma$ for $j=16$ and several values of $\alpha$. The curves display a pronounced dependence on the Lévy exponent, indicating that the interplay between nonlocal hops and resetting plays a crucial role in the encounter dynamics. For $\alpha=1$, corresponding to two normal random walkers, the MFET exhibits a clear minimum at $\gamma^\star = 0.0059$. This result indicates that when both walkers start at relatively short distances from the encounter node, with $d_{i_1,j}=15$ and $d_{i_2,j}=5$, the normal random walk with simultaneous resetting provides the most efficient strategy for both walkers to reach node $j=16$ simultaneously for the first time, minimizing the expected number of steps. As the motion of the second walker becomes more nonlocal, the optimal MFET increases and the variations near the minimum are less pronounced. Figure~\ref{Fig_5}(b) provides a two--dimensional representation of the MFET $\langle T(i_1,i_2; j,\gamma)\rangle$ as a function of the resetting probability 
$\gamma$ and the Lévy exponent $\alpha$ for the encounter node $j=16$. 
Encounter times are encoded by the color scale, allowing a direct visualization 
of the combined effect of resetting and the nonlocality of the second 
walker. The dashed curve indicates the value of $\gamma$ that minimizes the MFET 
for each $\alpha$, revealing how the optimal resetting probability varies with 
the Lévy exponent. 
\\[2mm]
In Figs.~\ref{Fig_5}(c)–(d), we repeat the same analysis for the encounter node $j=92$. In this case, the initial distance between the normal random walker and the encounter node is $d_{i_1,j}=10$, whereas the second walker, following the Lévy strategy, starts farther away, with $d_{i_2,j}=30$. For this configuration, the qualitative behavior of the MFET is reversed. As shown in Fig.~\ref{Fig_5}(c), the largest reduction of the MFET is obtained in the limit $\alpha \to 0$, where the Lévy walker performs long jumps with a high probability. The optimal resetting probability is $\gamma^\star = 0.0052$ in that case, whereas for $\alpha$ close to $1$ the MFET is significantly larger, although still with a pronounced dependence on $\gamma$. The two-dimensional representation of these curves in Fig.~\ref{Fig_5}(d) makes this contrast explicit. For $j=92$, the region of minimal encounter times is concentrated in the interval $0 < \alpha \leq 0.5$ and around the optimal resetting probability $\gamma^\star= 0.0052$, indicating that highly nonlocal dynamics combined with weak resetting provide the most efficient strategy for reducing encounter times. This behavior highlights the strong dependence of the optimal search strategy on both the location of the encounter node and the degree of nonlocality of the walkers.
\\[2mm]
Overall, the results shown in Fig.~\ref{Fig_5} demonstrate that simultaneous resetting interacts in a nontrivial manner with heterogeneous transport mechanisms. While resetting significantly enhances encounter efficiency for local dynamics and nearby targets, its effectiveness is reduced when combined with strongly nonlocal Lévy flights or when the encounter node lies far from the initial positions of the walkers. These findings also illustrate the flexibility of the theoretical framework developed in Sec.~\ref{Sec_GeneralTheory}, which naturally accommodates walkers with different transition rules and captures the interplay between resetting and long-range exploration.

\begin{figure*}[t!]
	\centering
	\includegraphics[width=0.99\linewidth]{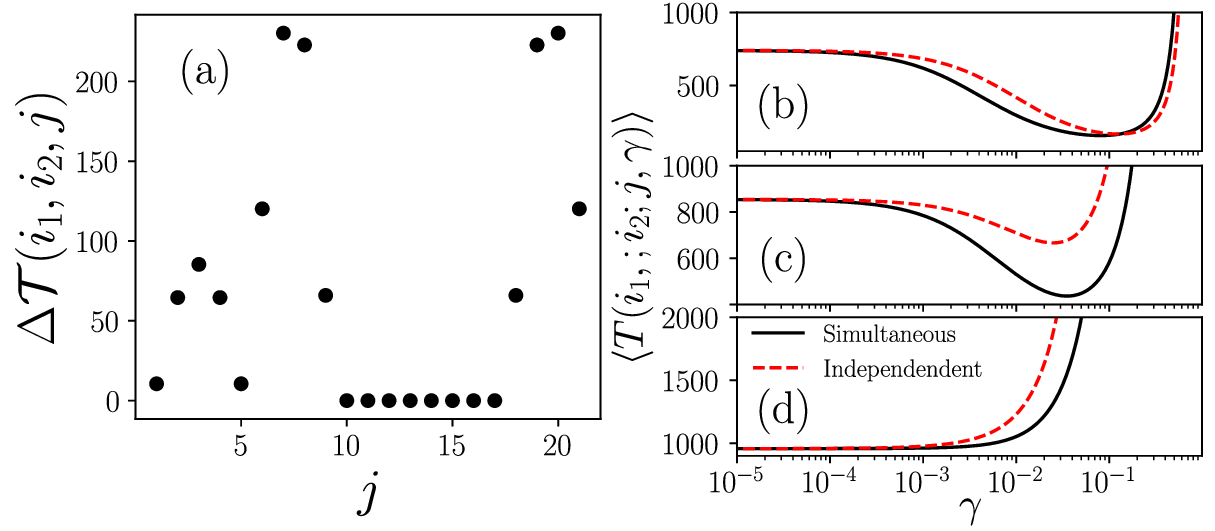}
	\\[-4mm]
	\caption{Comparison of the MFETs for two normal
		random walkers with simultaneous and independent resetting on a ring
		with $N=21$ nodes and initial positions $i_1=1$ and $i_2=5$.
		(a) $\Delta\mathcal{T}(i_1,i_2,j)$ [Eq.~(\ref{DeltaT_def})] as a
		function of the encounter node $j$. (b)–(d) MFET
		$\langle T(i_1,i_2,j,\gamma)\rangle$ as a function of the resetting
		probability $\gamma$ for $j=1$, $j=7$, and $j=10$, respectively.
		Solid lines correspond to simultaneous resetting, while dashed lines
		represent independent resetting with identical resetting probability.
		The ring topology and the initial positions of the walkers are shown
		in Fig.~\ref{Fig_2}(a).}
	\label{Fig_6}
\end{figure*}

\subsection{Comparing simultaneous and independent resetting protocols}
To quantify the effect of correlations between resetting events, we
compare the MFETs obtained for two
different resetting protocols. In the first case, which has
been explored analytically and numerically in the previous sections,
the walkers undergo
\emph{simultaneous resetting}, meaning that both particles return to
their initial positions at the same resetting times. 
In the second protocol, the walkers reset \emph{independently}: each
particle resets with the same probability $\gamma$, but the resetting
events are uncorrelated. This type of strategy has been analyzed in
Ref.~\cite{Rubio-Gomez_2025}, which studies multi-agent dynamics where
each random walker independently resets to its initial node.
\\[2mm]
To characterize the efficiency of each strategy, we focus on the
optimal MFET obtained by minimizing with respect to the resetting
probability $\gamma$. For two normal random walkers with simultaneous
resetting we define
\begin{equation}
	T^{(\mathrm{S})}_{\min}(i_1,i_2;j)=
	\min_{0\leq\gamma<1}
	\langle T(i_1,i_2;j,\gamma)\rangle_{\mathrm{S}},
\end{equation}
where $\langle T(i_1,i_2;j,\gamma)\rangle_{\mathrm{S}}$ denotes the
MFET for walkers starting at nodes $i_1$ and $i_2$ and encountering
each other at node $j$ under simultaneous resetting. Analogously, for
independent resetting events we define
\begin{equation}
	T^{(\mathrm{I})}_{\min}(i_1,i_2;j)=
	\min_{0\leq\gamma<1}
	\langle T(i_1,i_2;j,\gamma)\rangle_{\mathrm{I}},
\end{equation}
where $\langle T(i_1,i_2;j,\gamma)\rangle_{\mathrm{I}}$ denotes the
corresponding MFET (see Ref.~\cite{Rubio-Gomez_2025} for details).
\\[2mm]
The difference between the optimal efficiencies of the two protocols
is quantified through
\begin{equation}\label{DeltaT_def}
	\Delta\mathcal{T}(i_1,i_2,j)
	\equiv
	T^{(\mathrm{I})}_{\min}(i_1,i_2;j)
	-
	T^{(\mathrm{S})}_{\min}(i_1,i_2;j),
\end{equation}
which measures the advantage of simultaneous over independent
resetting.
\\[2mm]
Figure~\ref{Fig_6} compares both resetting protocols for two standard
random walkers moving on a ring with $N=21$ nodes and initial
positions $i_1=1$ and $i_2=5$. In Fig.~\ref{Fig_6}(a) we present
$\Delta\mathcal{T}(i_1,i_2,j)$ as a function of the encounter node $j$.
Figures~\ref{Fig_6}(b)–(d) show the dependence of the MFET
$\langle T(i_1,i_2,j,\gamma)\rangle$ on the resetting probability
$\gamma$ for representative encounter nodes $j$. Solid curves
correspond to simultaneous resetting, whereas dashed curves represent
independent resetting.
The comparison shows that, for all encounter nodes $j$,
$T^{(\mathrm{S})}_{\min}(i_1,i_2;j)\leq T^{(\mathrm{I})}_{\min}(i_1,i_2;j)$,
demonstrating that simultaneous resetting always leads to an optimal
MFET that is less than or equal to that obtained under independent
resetting. The extent of this advantage, however, depends on the
encounter node $j$. In particular, for nodes
$j=10,11,\ldots,17$,
$T^{(\mathrm{S})}_{\min}(i_1,i_2;j)=T^{(\mathrm{I})}_{\min}(i_1,i_2;j)$,
since neither protocol improves the encounter time relative to the
value obtained for $\gamma=0$, as illustrated in
Fig.~\ref{Fig_6}(d).
\\[2mm]
\begin{figure*}[t!]
	\centering
	\includegraphics[width=0.99\linewidth]{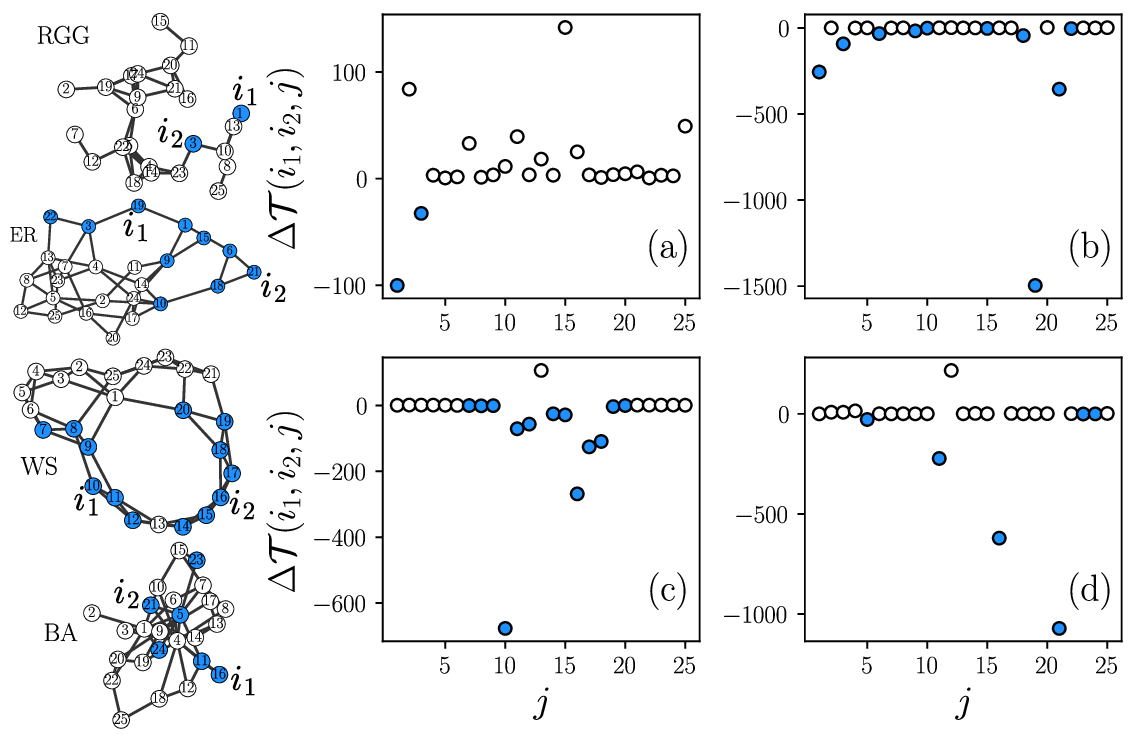}
	\\[-4mm]
	\caption{$\Delta\mathcal{T}(i_1,i_2,j)$ as a function of the encounter node $j$
		for two normal random walkers with simultaneous and independent
		resetting on heterogeneous networks with $N=25$ nodes.
		Panels correspond to (a) a random geometric graph (RGG),
		(b) an Erdős–Rényi (ER) network,
		(c) a Watts–Strogatz (WS) network, and
		(d) a Barabási–Albert (BA) network.
		The networks are shown on the left, with labeled nodes
		and initial positions $i_1$ and $i_2$ indicated.
		Node colors represent the sign of
		$\Delta\mathcal{T}(i_1,i_2,j)$: white nodes denote
		$\Delta\mathcal{T}\geq 0$ (simultaneous resetting favored or $\gamma=0$ is optimal), whereas
		blue nodes indicate $\Delta\mathcal{T}<0$ (independent resetting
		favored).}
	\label{Fig_7}
\end{figure*}
We now extend the analysis beyond regular rings and consider
heterogeneous network topologies. Figure~\ref{Fig_7} shows
$\Delta\mathcal{T}(i_1,i_2,j)$ as a function of the encounter node $j$
for the four network structures introduced in
Fig.~\ref{Fig_3}. The results reveal that the relative performance of the two resetting
protocols depends strongly on both the network topology and the
encounter node $j$. While simultaneous resetting is often more
efficient, heterogeneous structures give rise to situations in which
independent resetting becomes advantageous, as indicated by nodes with
$\Delta\mathcal{T}(i_1,i_2,j)<0$ (in blue). In these cases, the absence of
synchronization allows the walkers to explore the network more
independently, increasing the likelihood that one of them reaches the
encounter node without being forced to restart simultaneously with the
other. This effect is particularly relevant in heterogeneous networks,
where hubs or highly connected regions strongly influence the search
dynamics and the accessibility of specific nodes.
\section{Conclusions}
\label{Sec_Conclusions}
In this work, we have developed a general analytical framework to study first-encounter processes of multiple non-interacting random walkers evolving on networks under simultaneous stochastic resetting. Within this unified setting, we derived exact expressions for the mean first-encounter time (MFET) in terms of the spectral properties of the underlying dynamics without resetting, thereby extending previous results on single walkers and multi-agent encounters without resetting. We identify a general criterion to  determine when resetting is beneficial and when an optimal nonzero resetting probability exists. This criterion, expressed in terms of the first two moments of the encounter-time distribution of the process without resetting, are encapsulated in the coefficient $\Lambda$ introduced in Eq.~(\ref{Lambda_def}). Remarkably, this quantity depends solely on the spectral properties of the dynamics without resetting and provides a predictive measure to anticipate whether resetting reduces encounter times, independently of the resetting probability itself. 
\\[2mm]
Our analytical predictions were validated through a systematic investigation of representative network topologies. For two random walkers on rings and heterogeneous networks, we showed that the MFET strongly depends on the encounter node, with resetting leading to substantial reductions in encounter times only for specific targets. These effects persist in the presence of structural disorder and degree heterogeneity, demonstrating the robustness of the theoretical criterion across different classes of networks. We further showed that the phenomenology observed for two walkers naturally extends to encounters involving three and four walkers. In all cases, the sign of $\Lambda$ accurately predicts the existence of an optimal resetting probability. We also explored scenarios in which the walkers follow different underlying dynamics, focusing on the interplay between local random walks and nonlocal L\'evy flights. Our results show that resetting interacts non-trivially with long-range transport mechanisms, leading to strongly configuration-dependent optimal strategies. When both walkers start close to the encounter node, local dynamics combined with resetting yield the smallest mean first-encounter times. In contrast, when the walker performing L\'evy dynamics is far from the target, strongly nonlocal transport in the limit $\alpha\to0$ becomes favorable. These findings demonstrate that neither resetting nor nonlocal exploration is universally optimal, but that their efficiency is governed by the relative positions of the walkers and the encounter node. 
\\[2mm]
Additionally, by comparing simultaneous and independent resetting protocols for two normal random walkers, we showed that correlations between resetting events can significantly impact the optimal search efficiency. In particular, the advantages of each protocol depend strongly on the network structure, the initial conditions, and the encounter node, especially in heterogeneous networks.
\\[2mm]
This work provides a unified and flexible theoretical framework to analyze encounter times of multiple agents under simultaneous resetting on networks. Beyond its relevance to diffusive transport and stochastic search theory, our results may find applications in diverse contexts where coordinated encounters play a central role, including epidemic spreading, ecological interactions, human mobility, and distributed search and optimization problems. 

\appendix

\section{First-passage properties for multiple random walkers}
\label{Appendix_FPT}

In this Appendix, we calculate the first two moments of the first-encounter time $T(\vec{i},\vec{j})$, defined as the number of steps required for $S$ non-interacting random walkers under simultaneous resetting to reach for the first time the configuration $\vec{j}$, given the initial configuration $\vec{i}$ at $t=0$. For brevity in the notation, we have dropped the argument $\gamma$ of $T$, which is implicit. Our main goal is to obtain the mean first-encounter time (MFET) $\langle T(\vec{i},\vec{j})\rangle$ and the second moment $\langle T^2(\vec{i},\vec{j})\rangle$. The analytical formalism follows closely the theory of first-passage processes for a single random walker (see Refs. \cite{Hughes,NohRieger2004,reviewjcn_2021}), generalized to the configuration space $\mathcal{V}^S$. Our approach is similar to that developed in Ref. \cite{riascos_multiple2020} for synchronous random walkers, which we extend here to obtain explicit expressions for $\langle T^2(\vec{i},\vec{j})\rangle$.
\\[2mm]
The stochastic dynamics is governed by a Markov process defined by the transition matrix $\hat{\mathbf{\Pi}}$, introduced in Eq. (\ref{Def_Pi_matrix}), and satisfying the master equation (\ref{masterRrw}). Actually, the derivation here also holds for any discrete process with $S$ particles, not just those under simultaneous resetting. The occupation probability $\mathcal{P}(\vec{i},\vec{j};t)$ satisfies the renewal equation
\begin{equation}
	\label{EquF_PRE}
	\mathcal{P}(\vec{i},\vec{j};t)
	=
	\delta_{t,0}\delta_{\vec{i},\vec{j}}
	+
	\sum_{t'=0}^{t}
	F(\vec{i},\vec{j};t')
	\mathcal{P}(\vec{j},\vec{j};t-t'),
\end{equation}
where $F(\vec{i},\vec{j};t)$ denotes the first-passage probability to reach configuration $\vec{j}$ for the first time at step $t$. Introducing the discrete Laplace transform $\tilde{f}(\sigma)=\sum_{t=0}^{\infty}e^{-\sigma t}f(t)$, Eq.~(\ref{EquF_PRE}) gives
\begin{equation}
	\label{LaplTransF_PRE}
	\widetilde{F}(\vec{i},\vec{j};\sigma)
	=
	\frac{
		\widetilde{\mathcal{P}}(\vec{i},\vec{j};\sigma)
		-
		\delta_{\vec{i},\vec{j}}
	}{
		\widetilde{\mathcal{P}}(\vec{j},\vec{j};\sigma)
	}.
\end{equation}
The moments of the first-encounter time distribution follow from the expansion
\begin{equation}
	\widetilde{F}(\vec{i},\vec{j};\sigma)
	=
	1
	-
	\sigma\langle T(\vec{i},\vec{j})\rangle
	+
	\frac{\sigma^2}{2}
	\langle T^2(\vec{i},\vec{j})\rangle
	+
	O(\sigma^3).
\end{equation}
To compute these quantities, we introduce the $n$-th moments
\begin{equation}
	\label{Rij_n_PRE}
	\mathcal{R}^{(n)}(\vec{i},\vec{j})
	=
	\sum_{t=0}^{\infty}
	t^n
	\left[
	\mathcal{P}(\vec{i},\vec{j};t)
	-
	\mathcal{P}_{\vec{j}}^\infty(\vec{i})
	\right],
\end{equation}
where $\mathcal{P}_{\vec{j}}^\infty(\vec{i})$ denotes the stationary probability distribution in Eq. (\ref{Pinf_multiple}). The Laplace transform of the occupation probability admits the expansion
\begin{equation}
	\widetilde{\mathcal{P}}(\vec{i},\vec{j};\sigma)
	=
	\frac{\mathcal{P}_{\vec{j}}^\infty(\vec{i})}{1-e^{-\sigma}}
	+
	\sum_{n=0}^{\infty}
	(-1)^n
	\mathcal{R}^{(n)}(\vec{i},\vec{j})
	\frac{\sigma^n}{n!}.
\end{equation}
Substituting this expression into Eq.~(\ref{LaplTransF_PRE}) and expanding in powers of $\sigma$, the MFET is obtained as
\begin{equation}
	\label{MFPT_general_PRE}
	\langle T(\vec{i},\vec{j})\rangle
	=
	\frac{
		\mathcal{R}^{(0)}(\vec{j},\vec{j})
		-
		\mathcal{R}^{(0)}(\vec{i},\vec{j})
		+
		\delta_{\vec{i},\vec{j}}
	}{
		\mathcal{P}_{\vec{j}}^\infty(\vec{i})
	}.
\end{equation}
On the other hand, the transition probability admits the spectral decomposition
\begin{equation}\label{P_spectral_appendix}
	\mathcal{P}(\vec{i},\vec{j};t)
	=
	\sum_{\vec{l}\in\mathcal{V}^S}
	\zeta_{\vec{l}}^t
	\langle \vec{i}|\psi_{\vec{l}}\rangle
	\langle \bar{\psi}_{\vec{l}}|\vec{j}\rangle,
\end{equation}
where $\zeta_{\vec{l}}$ are the eigenvalues of $\hat{\mathbf{\Pi}}$ and
$|\psi_{\vec{l}}\rangle$, $\langle \bar{\psi}_{\vec{l}}|$
are the right and left eigenvectors, respectively.
\\[2mm]
To proceed, it is necessary to determine whether the largest eigenvalue $\zeta=1$ is degenerate. In the following, we assume that this eigenvalue is nondegenerate, as is the case for dynamics with simultaneous resetting, where $0<\gamma<1$. However, for $\gamma=0$, the eigenvalue $\zeta=1$ may be degenerate in bipartite networks (for example, rings with an even number of nodes and trees; see Ref. \cite{riascos_multiple2020} for a detailed discussion). Then, in terms of the set $\mathcal{I}=\mathcal{V}^S\setminus\{\vec{1}\}$ with $\vec{1}\equiv(1,1,\ldots,1)$, Eq.~(\ref{Rij_n_PRE}) gives
\begin{equation}
	\label{R0ij_PRE}
	\mathcal{R}^{(0)}(\vec{i},\vec{j})
	=
	\sum_{\vec{l}\in\mathcal{I}}
	\frac{
		\langle \vec{i}|\psi_{\vec{l}}\rangle
		\langle \bar{\psi}_{\vec{l}}|\vec{j}\rangle
	}{
		1-\zeta_{\vec{l}}
	}.
\end{equation}
Substituting into Eq.~(\ref{MFPT_general_PRE}), the MFPT for $\vec{i}\neq\vec{j}$ becomes
\begin{equation}
	\label{MFPT_spectral_PRE}
	\langle T(\vec{i},\vec{j})\rangle
	=
	\frac{1}{\mathcal{P}_{\vec{j}}^\infty(\vec{i})}
	\sum_{\vec{l}\in\mathcal{I}}
	\frac{
		\langle \vec{j}|\psi_{\vec{l}}\rangle
		\langle \bar{\psi}_{\vec{l}}|\vec{j}\rangle
		-
		\langle \vec{i}|\psi_{\vec{l}}\rangle
		\langle \bar{\psi}_{\vec{l}}|\vec{j}\rangle
	}{
		1-\zeta_{\vec{l}}
	}.
\end{equation}
On the other hand, the second moment of the first-passage time distribution is obtained from the second-order expansion of Eq.~(\ref{LaplTransF_PRE}), yielding
\begin{align}
	\label{T2ij_general_PRE}
	\langle T^2(\vec{i},\vec{j})\rangle
	&=
	\frac{
		(\mathcal{P}_{\vec{j}}^\infty(\vec{i})
		+
		2\mathcal{R}^{(0)}(\vec{j},\vec{j}))
		\delta_{\vec{i},\vec{j}}
	}{
		(\mathcal{P}_{\vec{j}}^\infty(\vec{i}))^2
	}
	\nonumber\\
	&+
	\frac{
		\mathcal{R}^{(0)}(\vec{j},\vec{j})
		-
		\mathcal{R}^{(0)}(\vec{i},\vec{j})
	}{
		\mathcal{P}_{\vec{j}}^\infty(\vec{i})
	}
	\nonumber\\
	&+
	2
	\frac{
		\mathcal{R}^{(1)}(\vec{j},\vec{j})
		-
		\mathcal{R}^{(1)}(\vec{i},\vec{j})
	}{
		\mathcal{P}_{\vec{j}}^\infty(\vec{i})
	}
	\nonumber\\
	&+
	2
	\frac{
		\mathcal{R}^{(0)}(\vec{j},\vec{j})
		\left[
		\mathcal{R}^{(0)}(\vec{j},\vec{j})
		-
		\mathcal{R}^{(0)}(\vec{i},\vec{j})
		\right]
	}{
		(\mathcal{P}_{\vec{j}}^\infty(\vec{i}))^2
	}.
\end{align}
Using the spectral decomposition in Eq. (\ref{P_spectral_appendix}), the quantity $\mathcal{R}^{(1)}(\vec{i},\vec{j})$ is given by
\begin{equation}
	\label{R1ij_PRE}
	\mathcal{R}^{(1)}(\vec{i},\vec{j})
	=
	\sum_{\vec{l}\in\mathcal{I}}
	\frac{
		\zeta_{\vec{l}}
	}{
		(1-\zeta_{\vec{l}})^2
	}
	\langle \vec{i}|\psi_{\vec{l}}\rangle
	\langle \bar{\psi}_{\vec{l}}|\vec{j}\rangle.
\end{equation}
Therefore, both the mean and the second moment of the first-passage time distribution are completely determined by the spectral properties of the transition matrix $\hat{\mathbf{\Pi}}$. These results constitute a direct generalization of the first-passage theory of a single random walker to the configuration space of $S$ non-interacting walkers, and provide a complete characterization of first-passage fluctuations in terms of the eigenvalues and eigenvectors of the underlying stochastic dynamics.

\section{Spectral properties of $\hat{\mathbf{\Pi}}(\vec{r};\gamma)$}\label{Appendix_Spec}
In this section, we derive the spectral properties of the transition operator describing $S$ independent random walkers subject to simultaneous stochastic resetting. The dynamics takes place in the configuration space $\mathcal{V}^S$. The transition operator is defined as a  combination of the collective random-walk operator $\hat{\mathcal W}$ and the resetting operator $\hat{\mathbf{\Theta}}(\vec r)$, which maps any configuration onto a fixed resetting configuration $\vec r=(r_1,r_2,\ldots,r_S)$. Our goal is to determine explicitly the eigenvalues and the associated left and right eigenvectors of this operator in terms of the spectral properties of the random-walk operator without resetting, $\hat{\mathcal W}$.
\subsubsection{General properties of $\hat{\mathbf{\Theta}}(\vec{r})$}
We consider the transition operator
\begin{equation}
	\hat{\mathbf{\Pi}}(\vec r;\gamma)
	=
	(1-\gamma)\hat{\mathcal W}
	+
	\gamma \hat{\mathbf{\Theta}}(\vec r),
	\label{Pi_def_multi_app}
\end{equation}
with $0 < \gamma < 1$. The configuration space is spanned by the basis vectors $|\vec i\rangle = 	|i_1\rangle \otimes |i_2\rangle \otimes \cdots \otimes |i_S\rangle$, where $i_s \in \{1,\dots,N\}$ denotes the position of walker $s$.
\\[2mm]
By definition, the resetting operator satisfies
\begin{equation}
	\langle \vec l|\hat{\mathbf{\Theta}}(\vec r)|\vec m\rangle
	=
	\delta_{\vec m,\vec r}
	=
	\prod_{s=1}^S \delta_{m_s,r_s},
	\label{Theta_matrix_multi}
\end{equation}
which shows that resetting projects any configuration onto $\vec r$.
\\[2mm]
On the other hand, since the walkers are independent, the collective transition operator for the dynamics without reset is \cite{riascos_multiple2020}
\begin{equation}
	\hat{\mathcal W}
	=
	\bigotimes_{s=1}^S \mathbf W^{(s)}.
\end{equation}
For ergodic random walks with $\mathbf W^{(s)}$ diagonalizable \cite{reviewjcn_2021}, its right and left eigenvectors satisfy
\begin{align}
	\mathbf W^{(s)}|\phi_{l_s}^{(s)}\rangle &= \lambda_{l_s}^{(s)} |\phi_{l_s}^{(s)}\rangle, \\
	\langle \bar{\phi}_{l_s}^{(s)}|\mathbf W^{(s)} &= \lambda_{l_s}^{(s)} \langle \bar{\phi}_{l_s}^{(s)}|.
\end{align}
Then, the eigenvectors of $\hat{\mathcal W}$ are given by the tensor products
\begin{equation}
	|\phi_{\vec l}\rangle =
	\bigotimes_{s=1}^S |\phi_{l_s}^{(s)}\rangle,
	\qquad
	\langle \bar{\phi}_{\vec l}| =
	\bigotimes_{s=1}^S \langle \bar{\phi}_{l_s}^{(s)}|,
\end{equation}
with eigenvalues
\begin{equation}
	\lambda_{\vec l}
	=
	\prod_{s=1}^S \lambda_{l_s}^{(s)}.
	\label{lambda_product_app}
\end{equation}
Furthermore, the eigenvectors of $\mathbf W^{(s)}$ satisfy the orthonormality and completeness relations 
\begin{align}
	\langle \bar{\phi}_{\vec l}|\phi_{\vec m}\rangle
	&=
	\delta_{\vec l,\vec m},
	\label{orth_phi_multi}
	\\
	\sum_{\vec l \in \mathcal{V}^S}
	|\phi_{\vec l}\rangle \langle \bar{\phi}_{\vec l}|
	&=
	\mathbf{I}^{\otimes S}.
	\label{complete_phi_multi}
\end{align}
The largest eigenvalue corresponds to $\vec 1=(1,1,\ldots,1)$ and satisfies $\lambda_{\vec 1}=1$. Since the corresponding left  eigenvector is uniform in configuration space, it follows that
\begin{equation}
	\sum_{\vec i \in \mathcal{V}^S}
	\langle \bar{\phi}_{\vec l }|\vec i\rangle
	=
	\frac{\delta_{\vec l,\vec 1}}
	{\langle \vec r|\phi_{\vec 1}\rangle}.
	\label{sum_identity_multi_app}
\end{equation}
Using the completeness relation in Eq.~(\ref{complete_phi_multi}) and   Eqs.~(\ref{Theta_matrix_multi}) and (\ref{sum_identity_multi_app}), the resetting operator can be written as
\begin{align}
	\hat{\mathbf{\Theta}}(\vec r)
	&= 	\sum_{\vec l,\vec m \in \mathcal{V}^S}
	|\phi_{\vec l}\rangle
	\langle \bar{\phi}_{\vec l}|
	\hat{\mathbf{\Theta}}(\vec r)
	|\phi_{\vec m}\rangle
	\langle \bar{\phi}_{\vec m}| 	\nonumber\\
	&=\sum_{\vec l,\vec m \in \mathcal{V}^S}\sum_{\vec i,\vec j \in \mathcal{V}^S} |\phi_{\vec l}\rangle
	\langle \bar{\phi}_{\vec l}|\vec{i}\rangle \delta_{\vec{j},\vec{r}} \langle \vec{j}	|\phi_{\vec m}\rangle
	\langle \bar{\phi}_{\vec m}| 
		\nonumber\\
	&=\sum_{\vec m \in \mathcal{V}^S} \langle \vec{r}	|\phi_{\vec m}\rangle \sum_{\vec l \in \mathcal{V}^S} |\phi_{\vec l}\rangle \sum_{\vec i \in \mathcal{V}^S}
	\langle \bar{\phi}_{\vec l}|\vec{i}\rangle 
	\langle \bar{\phi}_{\vec m}| 
	\nonumber\\
	&=\sum_{\vec m \in \mathcal{V}^S}
	\frac{\langle \vec r|\phi_{\vec m}\rangle}
	{\langle \vec r|\phi_{\vec 1}\rangle}
	|\phi_{\vec 1}\rangle
	\langle \bar{\phi}_{\vec m}|.
	\label{Theta_spec_multi_app}
\end{align}
This representation is useful for evaluating the action of $\hat{\mathbf{\Theta}}(\vec r)$ on the eigenvectors of the system.
\subsubsection{Eigenvalues and eigenvectors of $\hat{\mathbf{\Pi}}(\vec{r};\gamma)$}
We now determine the eigenvalues and eigenvectors of $\hat{\mathbf{\Pi}}(\vec r;\gamma)$. Acting on the $|\phi_{\vec 1}\rangle$ eigenvector yields
\begin{equation}
	\hat{\mathbf{\Pi}}(\vec r;\gamma)|\phi_{\vec 1}\rangle
	=
	|\phi_{\vec 1}\rangle,
\end{equation}
which shows that
\begin{equation}
	\zeta_{\vec 1}=1,
	\qquad
	|\psi_{\vec 1}\rangle=|\phi_{\vec 1}\rangle.
\end{equation}
In addition, for $\vec l \neq \vec 1$, we introduce the ansatz
\begin{equation}
	|\psi_{\vec l}\rangle
	=
	|\phi_{\vec l}\rangle
	+
	b_{\vec l}|\phi_{\vec 1}\rangle,
\end{equation}
which leads to the eigenvalues
\begin{equation}
	\zeta_{\vec l}
	=
	(1-\gamma)\lambda_{\vec l},
	\qquad \vec l \neq \vec 1,
\end{equation}
and right eigenvectors
\begin{equation}
	|\psi_{\vec l}\rangle
	=
	|\phi_{\vec l}\rangle
	-
	\frac{\gamma}{1-(1-\gamma)\lambda_{\vec l}}
	\frac{\langle \vec r|\phi_{\vec l}\rangle}
	{\langle \vec r|\phi_{\vec 1}\rangle}
	|\phi_{\vec 1}\rangle.
	\label{right_eigenvectors_multi_app}
\end{equation}
For the left eigenvectors, orthogonality implies
\begin{equation}\label{left_eigenvectors_lneq1_app}
	\langle \bar{\psi}_{\vec l}|
	=
	\langle \bar{\phi}_{\vec l}|,
	\qquad \vec l \neq \vec 1,
\end{equation}
while the left eigenvector $\langle \bar{\psi}_{\vec 1}|$ is given by
\begin{equation}
	\langle \bar{\psi}_{\vec 1}|
	=
	\langle \bar{\phi}_{\vec 1}|
	+
	\sum_{\vec m \in \mathcal{I}}
	\frac{\gamma}{1-(1-\gamma)\lambda_{\vec m}}
	\frac{\langle \vec r|\phi_{\vec m}\rangle}
	{\langle \vec r|\phi_{\vec 1}\rangle}
	\langle \bar{\phi}_{\vec m}|.
	\label{left_eigenvectors_multi_app}
\end{equation}
\subsubsection{Orthonormality and completeness of
	$\{|\psi_{\vec l}\rangle,\langle \bar{\psi}_{\vec l}|\}$}
Let us now discuss the orthonormality and completeness properties of the left and right eigenvectors of $\hat{\mathbf{\Pi}}(\vec r;\gamma)$. To establish orthonormality, we first consider the case $\vec l\neq\vec 1$ and $\vec m\neq\vec 1$. Using Eqs.~(\ref{right_eigenvectors_multi_app}) and (\ref{left_eigenvectors_lneq1_app}), we obtain
\begin{align}
	\langle \bar{\psi}_{\vec l}|\psi_{\vec m}\rangle
	&=
	\langle \bar{\phi}_{\vec l}|
	\left(
	|\phi_{\vec m}\rangle
	-
	\frac{\gamma}{1-(1-\gamma)\lambda_{\vec m}}
	\frac{\langle \vec r|\phi_{\vec m}\rangle}
	{\langle \vec r|\phi_{\vec 1}\rangle}
	|\phi_{\vec 1}\rangle
	\right)
	\nonumber\\
	&=
	\langle \bar{\phi}_{\vec l}|\phi_{\vec m}\rangle
	-
	\frac{\gamma}{1-(1-\gamma)\lambda_{\vec m}}
	\frac{\langle \vec r|\phi_{\vec m}\rangle}
	{\langle \vec r|\phi_{\vec 1}\rangle}
	\langle \bar{\phi}_{\vec l}|\phi_{\vec 1}\rangle
	\nonumber\\
	& =
	\delta_{\vec l,\vec m},
	\qquad
	\vec l,\vec m\neq \vec 1.
\end{align}
We now consider $\vec l\neq \vec 1$ and $\vec m=\vec 1$. Using
$|\psi_{\vec 1}\rangle=|\phi_{\vec 1}\rangle$, we obtain
\begin{equation}
	\langle \bar{\psi}_{\vec l}|\psi_{\vec 1}\rangle
	=
	\langle \bar{\phi}_{\vec l}|\phi_{\vec 1}\rangle
	=
	0,
	\qquad
	\vec l\neq \vec 1.
\end{equation}
Similarly, for $\vec l=\vec 1$ and $\vec m\neq \vec 1$, using
Eq.~(\ref{left_eigenvectors_multi_app}), we obtain
\begin{equation*}
	\langle \bar{\psi}_{\vec 1}|\psi_{\vec m}\rangle
	=
	\left(
	\langle \bar{\phi}_{\vec 1}|
	+
	\sum_{\vec k\in \mathcal{I}}
	\frac{\gamma}{1-(1-\gamma)\lambda_{\vec k}}
	\frac{\langle \vec r|\phi_{\vec k}\rangle}
	{\langle \vec r|\phi_{\vec 1}\rangle}
	\langle \bar{\phi}_{\vec k}|
	\right)
	|\psi_{\vec m}\rangle.
\end{equation*}
Using Eq.~(\ref{right_eigenvectors_multi_app}), we obtain two contributions,
\begin{align}
	\langle \bar{\phi}_{\vec 1}|\psi_{\vec m}\rangle
	&=
	-\frac{\gamma}{1-(1-\gamma)\lambda_{\vec m}}
	\frac{\langle \vec r|\phi_{\vec m}\rangle}
	{\langle \vec r|\phi_{\vec 1}\rangle},
	\\
	\langle \bar{\phi}_{\vec k}|\psi_{\vec m}\rangle
	&=
	\delta_{\vec k,\vec m}.
\end{align}
Therefore,
\begin{multline}
	\langle \bar{\psi}_{\vec 1}|\psi_{\vec m}\rangle
	=
	-\frac{\gamma}{1-(1-\gamma)\lambda_{\vec m}}
	\frac{\langle \vec r|\phi_{\vec m}\rangle}
	{\langle \vec r|\phi_{\vec 1}\rangle}
	\\+
	\frac{\gamma}{1-(1-\gamma)\lambda_{\vec m}}
	\frac{\langle \vec r|\phi_{\vec m}\rangle}
	{\langle \vec r|\phi_{\vec 1}\rangle}
	=
	0,
	\qquad
	\vec m\neq \vec 1.
\end{multline}
Finally, for $\vec l=\vec 1$ and $\vec m = \vec 1$,
\begin{equation}
	\langle \bar{\psi}_{\vec 1}|\psi_{\vec 1}\rangle
	=
	\langle \bar{\psi}_{\vec 1}|\phi_{\vec 1}\rangle
	=
	\langle \bar{\phi}_{\vec 1}|\phi_{\vec 1}\rangle
	=
	1.
\end{equation}
Combining all cases, we conclude that
\begin{equation}
	\langle \bar{\psi}_{\vec l}|\psi_{\vec m}\rangle
	=
	\delta_{\vec l,\vec m}.
\end{equation}
We now prove completeness. Using the definitions of the eigenvectors, we write
\begin{align}
	\sum_{\vec l \in \mathcal{V}^S}
	|\psi_{\vec l}\rangle
	\langle \bar{\psi}_{\vec l}|
	&=
	|\psi_{\vec 1}\rangle
	\langle \bar{\psi}_{\vec 1}|
	+
	\sum_{\vec l\in \mathcal{I}}
	|\psi_{\vec l}\rangle
	\langle \bar{\phi}_{\vec l}|.
\end{align}
Substituting Eq.~(\ref{right_eigenvectors_multi_app}),
\begin{multline}
	\sum_{\vec l\in \mathcal{I}}
	|\psi_{\vec l}\rangle
	\langle \bar{\phi}_{\vec l}|
	=
	\sum_{\vec l\in \mathcal{I}}
	|\phi_{\vec l}\rangle
	\langle \bar{\phi}_{\vec l}|
	\\ -
	|\phi_{\vec 1}\rangle
	\sum_{\vec l\in \mathcal{I}}
	\frac{\gamma}{1-(1-\gamma)\lambda_{\vec l}}
	\frac{\langle \vec r|\phi_{\vec l}\rangle}
	{\langle \vec r|\phi_{\vec 1}\rangle}
	\langle \bar{\phi}_{\vec l}|.
\end{multline}
Using Eq.~(\ref{left_eigenvectors_multi_app}), we recognize
\begin{equation}
	\sum_{\vec l\in \mathcal{I}}
	\frac{\gamma}{1-(1-\gamma)\lambda_{\vec l}}
	\frac{\langle \vec r|\phi_{\vec l}\rangle}
	{\langle \vec r|\phi_{\vec 1}\rangle}
	\langle \bar{\phi}_{\vec l}|
	=
	\langle \bar{\psi}_{\vec 1}|
	-
	\langle \bar{\phi}_{\vec 1}|.
\end{equation}
Therefore,
\begin{align}
	\sum_{\vec l \in \mathcal{V}^S}
	|\psi_{\vec l}\rangle
	\langle \bar{\psi}_{\vec l}|
	&=
	|\phi_{\vec 1}\rangle
	\langle \bar{\psi}_{\vec 1}|
	+
	\sum_{\vec l\in \mathcal{I}}
	|\phi_{\vec l}\rangle
	\langle \bar{\phi}_{\vec l}|
	-
	|\phi_{\vec 1}\rangle
	\left(
	\langle \bar{\psi}_{\vec 1}|
	-
	\langle \bar{\phi}_{\vec 1}|
	\right)
	\nonumber\\
	&=
	\sum_{\vec l \in \mathcal{V}^S}
	|\phi_{\vec l}\rangle
	\langle \bar{\phi}_{\vec l}|.
\end{align}
Finally, using the completeness relation (\ref{complete_phi_multi}), we obtain
\begin{equation}
	\sum_{\vec l \in \mathcal{V}^S}
	|\psi_{\vec l}\rangle
	\langle \bar{\psi}_{\vec l}|
	=
	\mathbf{I}^{\otimes S}.
\end{equation}
Thus, the eigenvectors of $\hat{\mathbf{\Pi}}(\vec r;\gamma)$ form a complete orthonormal basis in the configuration space. These spectral properties provide the foundation for the analytical calculation of dynamical observables, including occupation probabilities and mean first-encounter times.
\section{When resetting becomes advantageous?}\label{Appendix_Adv}
Let us now to consider the mean first–encounter time (MFET) for dynamics with stochastic resetting to the initial configuration, $\vec r=\vec i$. Using the spectral representation derived in Sec.\ref{Appendix_FPT} and the eigenvalues
$\zeta_{\vec l}=(1-\gamma)\lambda_{\vec l}$, the MFET $\langle T(\vec i,\vec j,\gamma)\rangle$ for $\vec i\neq\vec j$ can be written as Eq. (\ref{MFPT_resetSM}), where $\vec{j}=(j,j,\ldots,j)$, with $j$ denoting the node at which all walkers coincide for the first time. Following the same approach as in the single-walker case in Ref. \cite{riascos_jpa_2022}, it is convenient to introduce the spectral sums
\begin{align}
	\mathcal C_{\vec i,\vec j}(\gamma)
	&\equiv
	\sum_{\vec l\in\mathcal I}
	\frac{
		\langle \vec j|\phi_{\vec l}\rangle
		\langle\bar\phi_{\vec l}|\vec j\rangle
		-
		\langle \vec i|\phi_{\vec l}\rangle
		\langle\bar\phi_{\vec l}|\vec j\rangle
	}
	{1-(1-\gamma)\lambda_{\vec l}},
	\\
	\mathcal S_{\vec i,\vec j}(\gamma)
	&\equiv
	\sum_{\vec l\in\mathcal I}
	\frac{
		\langle \vec i|\phi_{\vec l}\rangle
		\langle\bar\phi_{\vec l}|\vec j\rangle
	}
	{1-(1-\gamma)\lambda_{\vec l}}, \label{Sij_definition}
\end{align}
so that Eq.~(\ref{MFPT_resetSM}) becomes
\begin{equation}
	\langle T(\vec i,\vec j,\gamma)\rangle
	=
	\frac{\mathcal C_{\vec i,\vec j}(\gamma)}
	{P^\infty_{\vec j}(\vec i,0)+\gamma\,\mathcal S_{\vec i,\vec j}(\gamma)} .
	\label{MFET_compact_multi_consistent}
\end{equation}
Then, the optimal resetting probability $\gamma^\star$ minimizes the MFET and satisfies
\begin{equation}
	\frac{d}{d\gamma}
	\langle T(\vec i,\vec j,\gamma)\rangle
	\Big|_{\gamma=\gamma^\star}
	=0 .
	\label{opt_condition_multi_consistent}
\end{equation}
Differentiating Eq.~(\ref{MFET_compact_multi_consistent}) yields
\begin{equation}
	\mathcal C'_{\vec i,\vec j}(\gamma^\star)
	=
	\langle T(\vec i,\vec j,\gamma^\star)\rangle
	\left[
	\gamma^\star \mathcal S'_{\vec i,\vec j}(\gamma^\star)
	+
	\mathcal S_{\vec i,\vec j}(\gamma^\star)
	\right],
	\label{opt_condition_C_multi_consistent}
\end{equation}
where $\mathcal{F}'(\gamma^\star)$ denotes $\frac{d}{d\gamma}\mathcal{F}(\gamma)\Big|_{\gamma=\gamma^\star}$. 
\\[2mm]
We now express this relation in terms of the moments $\mathcal{R}^{(n)}(\vec i,\vec j)$ defined in Eq.~(\ref{Rij_n_PRE}). Differentiating $\mathcal C_{\vec i,\vec j}(\gamma)$ gives
\begin{align}
	\mathcal C'_{\vec i,\vec j}(\gamma)
	&=
	-
	\sum_{\vec l\in\mathcal I}
	\frac{\lambda_{\vec l}}
	{\left[1-(1-\gamma)\lambda_{\vec l}\right]^2}
	\nonumber\\
	&\quad\times
	\left[
	\langle \vec j|\phi_{\vec l}\rangle
	\langle\bar\phi_{\vec l}|\vec j\rangle
	-
	\langle \vec i|\phi_{\vec l}\rangle
	\langle\bar\phi_{\vec l}|\vec j\rangle
	\right].
\end{align}
On the other hand, using Eq.~(\ref{R1ij_PRE}) and $\zeta_{\vec l}=(1-\gamma)\lambda_{\vec l}$, we obtain
\begin{multline}
	\mathcal R^{(1)}(\vec j,\vec j)
	-
	\mathcal R^{(1)}(\vec i,\vec j)=
	\\
	\sum_{\vec l\in\mathcal I}
	\frac{(1-\gamma)\lambda_{\vec l}}
	{\left[1-(1-\gamma)\lambda_{\vec l}\right]^2}
	\left[
	\langle \vec j|\phi_{\vec l}\rangle
	\langle\bar\phi_{\vec l}|\vec j\rangle
	-
	\langle \vec i|\phi_{\vec l}\rangle
	\langle\bar\phi_{\vec l}|\vec j\rangle
	\right].
\end{multline}
Comparing both expressions, we obtain the exact relation
\begin{equation}
	\mathcal C'_{\vec i,\vec j}(\gamma)
	=
	-\frac{1}{1-\gamma}
	\left[
	\mathcal R^{(1)}(\vec j,\vec j)
	-
	\mathcal R^{(1)}(\vec i,\vec j)
	\right].
	\label{Cprime_R1_consistent}
\end{equation}
Substituting into Eq.~(\ref{opt_condition_C_multi_consistent}) yields
\begin{multline}
	\mathcal R^{(1)}(\vec j,\vec j)
	-
	\mathcal R^{(1)}(\vec i,\vec j)=
	\\
	-(1-\gamma^\star)
	\langle T(\vec i,\vec j;\gamma^\star)\rangle
	\left[
	\gamma^\star \mathcal S'_{\vec i,\vec j}(\gamma^\star)
	+
	\mathcal S_{\vec i,\vec j}(\gamma^\star)
	\right].
	\label{R1_relation_consistent}
\end{multline}
%
Using the general expression for the second moment in Eq.~(\ref{T2ij_general_PRE}), for $\vec i\neq \vec j$, this expression can be rearranged as
\begin{multline}
	\langle T^2(\vec i,\vec j,\gamma)\rangle
	=
	\langle T(\vec i,\vec j,\gamma)\rangle
	+
	2
	\frac{
		\mathcal{R}^{(1)}(\vec j,\vec j)
		-
		\mathcal{R}^{(1)}(\vec i,\vec j)
	}{
		\mathcal{P}^{\infty}_{\vec{j}}(\vec{i},\gamma)
	}
	\\+
	2
	\frac{
		\mathcal{R}^{(0)}(\vec j,\vec j)
	}{
		\mathcal{P}^{\infty}_{\vec{i}}(\vec{r},\gamma)
	}
	\langle T(\vec i,\vec j,\gamma)\rangle .
	\label{T2_expand_multi_PRE2}
\end{multline}
On the other hand, from the spectral definition of $\mathcal S_{\vec i,\vec j}(\gamma)$ in Eq.~(\ref{Sij_definition}), one finds for $\vec i\neq \vec j$,
\begin{equation}
	(1-\gamma)
	\left[
	\gamma \mathcal S'_{\vec i,\vec j}(\gamma)
	+
	\mathcal S_{\vec i,\vec j}(\gamma)
	\right]
	=
	\mathcal{R}^{(0)}(\vec i,\vec j).
	\label{S_identity_PRE}
\end{equation}
Evaluating Eq.~(\ref{S_identity_PRE}) at $\gamma^\star$ and substituting into Eq.~(\ref{R1_relation_consistent}) yields
\begin{equation}
	\mathcal{R}^{(1)}(\vec j,\vec j)
	-
	\mathcal{R}^{(1)}(\vec i,\vec j)
	=
	-
	\mathcal{R}^{(0)}(\vec i,\vec j)
	\langle T(\vec i,\vec j,\gamma^\star)\rangle .
	\label{R1_final_PRE}
\end{equation}
Substituting this result into Eq.~(\ref{T2_expand_multi_PRE2}) gives
\begin{multline}
	\langle T^2(\vec i,\vec j,\gamma^\star)\rangle
	=
	\langle T(\vec i,\vec j,\gamma^\star)\rangle
	\\ +
	2
	\frac{
		\mathcal{R}^{(0)}(\vec j,\vec j)
		-
		\mathcal{R}^{(0)}(\vec i,\vec j)
	}{
	\mathcal{P}^{\infty}_{\vec{j}}(\vec{i},\gamma)
	}
	\langle T(\vec i,\vec j,\gamma^\star)\rangle .
\end{multline}
Finally, using Eq.~(\ref{MFPT_general_PRE}), we obtain
\begin{equation}
	\langle T^2(\vec i,\vec j,\gamma^\star)\rangle
	=
	\langle T(\vec i,\vec j,\gamma^\star)\rangle
	+
	2\,\langle T(\vec i,\vec j,\gamma^\star)\rangle^2,
	\label{T2_opt_consistent}
\end{equation}
which provides an exact relation between the first two moments of the first–passage time distribution at optimal resetting. 
Introducing the coefficient of variation defined in Eq.~(\ref{z_definition}), we obtain the fluctuation criterion at optimal resetting,
 \begin{equation}
	z^2(\vec i,\vec j,\gamma^\star)
	=
	1+
	\frac{1}{\langle T(\vec i,\vec j,\gamma^\star)\rangle},
	\qquad
	\vec i\neq\vec j .
	\label{z_opt_consistent}
\end{equation}
Similarly, resetting improves the search if introducing an infinitesimal resetting probability reduces the MFET. This condition can be written as
\begin{equation}
	\left.
	\frac{d}{d\gamma}
	\langle T(\vec i,\vec j,\gamma)\rangle
	\right|_{\gamma\to0}
	<0,
	\qquad
	\vec i\neq\vec j
	\label{resetting_improves_condition}
\end{equation}
and, considering Eq.~(\ref{MFET_compact_multi_consistent}), we differentiate with respect to $\gamma$ and evaluate at $\gamma=0$ to obtain
\begin{align}
	\left.
	\frac{d}{d\gamma}
		\langle T(\vec i,\vec j,\gamma)\rangle
	\right|_{\gamma=0}
	&=
	\frac{
		\mathcal C'_{\vec i,\vec j}(0)
		P^\infty_{\vec j}(\vec i,0)
		-
		\mathcal C_{\vec i,\vec j}(0)
		\mathcal S_{\vec i,\vec j}(0)
	}
	{\left[P^\infty_{\vec j}(\vec i,0)\right]^2}.
	\label{dTdgamma_general}
\end{align}
Using Eq.~(\ref{MFET_compact_multi_consistent}), evaluated at $\gamma=0$, for $\vec i\neq\vec j$
\begin{equation}
	\langle T(\vec i,\vec j,0)\rangle
	=
	\frac{\mathcal C_{\vec i,\vec j}(0)}
	{P^\infty_{\vec j}(\vec i,0)},
	\label{MFPT_noreset}
\end{equation}
we rewrite Eq.~(\ref{dTdgamma_general}) as
\begin{multline}
	\left.
	\frac{d}{d\gamma}
		\langle T(\vec i,\vec j,\gamma)\rangle
	\right|_{\gamma=0}
	=\\
	\frac{1}{P^\infty_{\vec j}(\vec i,0)}
	\left[
	\mathcal C'_{\vec i,\vec j}(0)
	-
	\langle T(\vec i,\vec j,0)\rangle
	\mathcal S_{\vec i,\vec j}(0)
	\right].
	\label{dTdgamma_simplified}
\end{multline}
Using the exact relation obtained previously, Eq.~(\ref{Cprime_R1_consistent}),
\begin{equation}
	\mathcal C'_{\vec i,\vec j}(0)
	=
	-
	\left[
	\mathcal R^{(1)}(\vec j,\vec j)
	-
	\mathcal R^{(1)}(\vec i,\vec j)
	\right],
	\label{Cprime_gamma0}
\end{equation}
and recalling from the definition of $\mathcal R^{(0)}(\vec i,\vec j)$ that at $\gamma=0$
\begin{equation}
	\mathcal S_{\vec i,\vec j}(0)
	=
	\mathcal R^{(0)}(\vec i,\vec j),
	\label{S_gamma0}
\end{equation}
we obtain
\begin{multline}
	\left.
	\frac{d}{d\gamma}
		\langle T(\vec i,\vec j,\gamma)\rangle
	\right|_{\gamma=0}
	=
	-
	\frac{
		\mathcal R^{(1)}(\vec j,\vec j)
		-
		\mathcal R^{(1)}(\vec i,\vec j)
	}
	{P^\infty_{\vec j}(\vec i,0)}
	\\
	-
	\frac{
		\mathcal R^{(0)}(\vec i,\vec j)
	}
	{P^\infty_{\vec j}(\vec i;0)}
	\langle T(\vec i,\vec j,0)\rangle.
	\label{dTdgamma_Rmoments}
\end{multline}
Using Eq.~(\ref{T2ij_general_PRE}) to eliminate $\mathcal R^{(1)}(\vec j,\vec j)-\mathcal R^{(1)}(\vec i,\vec j)$, we finally obtain for $\vec i\neq\vec j$
\begin{multline}
	\left.
	\frac{d}{d\gamma}
		\langle T(\vec i,\vec j,\gamma)\rangle
	\right|_{\gamma=0}
	=\\
	\frac{
		\langle T^2(\vec i,\vec j,0)\rangle
		-
		\langle T(\vec i,\vec j,0)\rangle
		-
		2\langle T(\vec i,\vec j,0)\rangle^2
	}
	{2\langle T(\vec i,\vec j,0)\rangle}.
	\label{dTdgamma_variance_form}
\end{multline}
Finally, introducing the coefficient of variation defined in Eq.~(\ref{z_definition}), the condition (\ref{resetting_improves_condition}) becomes
\begin{equation}
	z^2(\vec i,\vec j,0)
	>
	1+
	\frac{1}{\langle T(\vec i,\vec j,0)\rangle},
	\qquad
	\vec i\neq\vec j.
	\label{z_smallgamma_consistent}
\end{equation}
This result establishes that stochastic resetting accelerates the encounter dynamics when the relative fluctuations of the first–encounter time without resetting exceed a well-defined threshold determined by the mean encounter time.


\end{document}